\documentclass[10pt,a4paper]{article}
\usepackage[cp1250]{inputenc}
\usepackage[T1]{fontenc}
\usepackage{amsmath}
\usepackage{amsfonts}
\usepackage{amssymb}
\usepackage{amsthm}
\usepackage{graphicx}
\usepackage{algorithm}
\usepackage{algcompatible}
\usepackage{algpseudocode}
\usepackage{caption}
\usepackage{subfigure}
\usepackage{color}   % colour of a text
\usepackage{hyperref}
\usepackage{breakcites}   % citations break at the end of margin
\usepackage[left=1in,top=1in,right=1in,bottom=1in,nohead,paperwidth=8.5in, paperheight=11in]{geometry} 

\usepackage{natbib}

\theoremstyle{definition}
\newtheorem{definition}{Definition}[section]

\title{\bf
	Principal Balances of Compositional Data for Regression and Classification using Partial Least Squares}
\vspace{3mm}
\author{V. Nesrstov\'a$^{1}$, I. Wilms$^{2}$, J. Palarea-Albaladejo$^{3}$, P. Filzmoser$^{4}$, J.A. Mart\'in-Fern\'andez$^{3}$, \\ D. Friedeck\'y$^{5}$ and K. Hron$^{1}$}

\date{}  
\begin{document}
	\maketitle
	\noindent
	{\small
		$^{1}$Department of Mathematical Analysis and Applications of Mathematics, Palack\'{y} University Olomouc, Faculty of Science, 17. listopadu 12, Olomouc, Czech Republic; \textit{viktorie.nesrstova@gmail.cz} \\
		$^{2}$Department of Quantitative Economics, Maastricht University, Tongersestraat 53, Maastricht, The Netherlands\\
		$^{3}$Department of Computer Science, Applied Mathematics and Statistics, University of Girona, Campus Montilivi, Edifici P-4, 17003 Girona, Spain\\
		$^{4}$Department of Statistics and Probability Theory, Vienna University of Technology, Wiedner Hauptstrasse 8-10, Vienna, Austria \\
		$^{5}$Laboratory for Inherited Metabolic Disorders, Department of Clinical Biochemistry, University Hospital Olomouc and Faculty of Medicine and Dentistry, Palack\'{y} University Olomouc, I. P. Pavlova 6, 779 00 Olomouc, Czech Republic}

	\bigskip
	
	\begin{abstract}
		High-dimensional compositional data are commonplace in the modern omics sciences amongst others. Analysis of compositional data requires a proper choice of orthonormal coordinate representation as their relative nature is not compatible with the direct use of standard statistical methods. Principal balances, a specific class of log-ratio coordinates, are well suited to this context since they are constructed in such a way that the first few coordinates capture most of the variability in the original data. Focusing on regression and classification problems in high dimensions, we propose a novel Partial Least Squares (PLS) based procedure to construct principal balances that maximize explained variability of the response variable and notably facilitates interpretability when compared to the ordinary PLS formulation. The proposed PLS principal balance approach can be understood as a generalized version of common logcontrast models, since multiple orthonormal (instead of one) logcontrasts are estimated simultaneously. We demonstrate the performance of the method using both simulated and real data sets.
		\bigskip
		
		\noindent \textbf{Keywords: }{Compositional data, Balance coordinates, PLS regression and classification, High-dimensional data, Metabolomic data}
		
	\end{abstract}
	
	\newtheorem{defi}{Definition} [section]   % define style of a DEF (cursive)

	\newpage
	%%%%%%%%%%%%%%%%%%%%%%%%%%%%%%%%%%%%%%%%%%%%%%%%%%%%%%%%%%%%%%%%%%%%%%%%
	\section{Introduction}  \label{Int}
	%%%%%%%%%%%%%%%%%%%%%%%%%%%%%%%%%%%%%%%%%%%%%%%%%%%%%%%%%%%%%%%%%%%%%%%%
	
	Compositional data (CoDa) occur in plenty of research fields, such as geochemistry \citep{hron2021analysing}, metabolomics \citep{stefelova2021weighted}, microbiome data \citep{monti2021robust}, time use data \citep{Dum+2020}, or ecology \citep{Per+2020}. Let us consider a regression (or classification) task where  $\mathbf{y} = (y_{1},\dots,y_{n})^\top$ is a vector of $n$ observations of a continuous (or binary) response variable and  $\mathbf{X}=(x_{ij})_{1\leq i \leq n, 1\leq j \leq D}$ is an associated matrix of a $D$-part compositional predictor. Analyzing CoDa requires careful consideration since such data do not carry relevant information in their absolute values, but rather in the ratios between the parts that constitute the composition. Moreover, in this paper we address the case of high-dimensional compositions consisting of a large number $D$ of parts.
	
	In the growing literature regarding omics sciences, most data are actually of relative (hence compositional) nature \citep{gloor17} and the development of methods for high-dimensional compositions is of increasing interest. The challenge not only concerns compositions consisting of many parts, but also the fact that the number of samples $n$ is usually substantially smaller than the number $D$ of parts due to the nature of the technology and omics sciences in general (as for example hundreds of proteins or metabolites are examined). Such settings with more variables than observations immediately discard the use of the most popular regression and classification methods, including Least Squares (LS) regression and Linear or Quadratic Discriminant Analysis (LDA/QDA) models. 
	
	For regression analysis with high-dimensional compositions, logcontrast models \citep{aitchison84} have gained increasing popularity, see for instance \cite{bates2019log}, \cite{susin2020variable} and \cite{gordon2022learning}. Logconstrasts are a building block in CoDa analysis through the log-ratio methodology \citep{aitchison1986statistical}. Given a $D$-part composition, a logconstrast is a loglinear combination 
	$$
	\sum^{D}_{i=1} a_{i}\ln x_{i}, \quad \textrm{with} \quad \sum^{D}_{i=1} a_{i} = 0, \quad a_{i} \in \mathbb{R}.
	$$
	Any log-ratio coordinate representation of CoDa consists of $D-1$ logcontrasts, with such number corresponding to the actual dimensionality of compositions. A typical feature of logcontrast models in \cite{aitchison84} or \cite{rivera2018balances} is that only one of the possible $D-1$ logcontrasts corresponding to the dimensionality of $D$-part CoDa is estimated as predictor variable. Note that considering only one logconstrast might be unnecessarily restrictive as other logcontrasts can be of interest as predictor variables, this is something we investigate here.
	
	However, estimating the set of all possible $D-1$ predictor logcontrasts could be computationally exhaustive or even unfeasible with high-dimensional compositions, definitely so when using existing logcontrast models. But, in fact, this would not be needed. Just having a few logcontrasts capturing most of the information contained in the original explanatory composition, while they appropriately relate to the response variable, would be necessary. We here introduce a model that sufficiently explains a response variable $\mathbf{y}$ using the matrix $\mathbf{X}$ of compositional predictors while performing dimension reduction through Partial Least Squares (PLS) regression/classification \citep{wold2001pls}. Prior to PLS modeling it is necessary to express CoDa in a proper log-ratio coordinate system. For example, a common choice is to use clr coefficients \citep{gallo10} and subsequently perform PLS analysis on them, although alternative coordinate representations can be considered \citep{kalivodova2015pls,stefelova2021weighted}. Note that clr coefficients have a direct link to logcontrasts \citep{martin2018advances}.
	
	Even though PLS is a convenient method to model relationships between response and explanatory variables, when dealing with compositions, the interpretation of logcontrasts might become challenging in high dimensions and thus simplification would be welcome. To this end, we propose to use a special class of log-ratio coordinates, so-called balance coordinates \citep{egozcue05}, which are interpretable in terms of contrasts between subgroups of compositional parts summarized by their geometric means. Note also that balances aggregate all pairwise log-ratios between parts with positive and negative sign in the respective logcontrast \citep{hron18}. Furthermore, in order to achieve orthonormality between balance coordinates, we tailor the original principal balances (PB) approach of \cite{pawlowsky2011principal} and \citep{martin2018advances}. The original PB method was developed to enhance interpretability in dimension reduction of CoDa in the style of principal component analysis (we will denote this PCA-PB). We adapt it here to involve a response variable within a regression or classification problem by replacing PCA by PLS. Accordingly, up to $D-1$ PBs with decreasing explanatory power are obtained that can be either directly interpreted or used for further statistical analysis. In the following, we will refer to this new proposal as \textit{PLS-PB}.
	
	This manuscript is structured as follows. In Section \ref{Method}, the basics of CoDa are presented together with the description of the proposed PLS-PB procedure. In Section \ref{sec:simulations}, a simulation study is conducted to investigate the ability of PLS-PB to reflect and simplify the structure of PLS loadings as well as its prediction performance. The method is then applied to two real-world data sets in Section \ref{RealDat}. Section \ref{end} concludes with some final remarks and future outlook.

	%%%%%%%%%%%%%%%%%%%%%%%%%%%%%%%%%%%%%%%%%%%%%%%%%%%%%%%%%%%%%%%%%%%%%%%%
	\section{Compositional Data and PLS Principal Balances}   \label{Method}
	%%%%%%%%%%%%%%%%%%%%%%%%%%%%%%%%%%%%%%%%%%%%%%%%%%%%%%%%%%%%%%%%%%%%%%%%
	When analyzing CoDa, their relative nature needs to be appropriately accounted for. The sample space of CoDa is formed by equivalence classes of proportional vectors \citep{BM2016}. In practice, CoDa are typically (equivalently) represented in the form of percentages, proportions or parts per million (ppm), that is, as data with a constant sum constraint and living on a simplex. Given the scale invariance property of compositions \citep{aitchison1986statistical} and the geometric structure of their sample space (the so-called Aitchison geometry), a well-principled way to conduct analysis of CoDa is to express them in the form of log-ratios of parts, or more generically as their logcontrasts, and then proceed to further statistical processing on these \citep{pawlowsky2015modeling,filzmoser18}.
	
	Log-contrast models have been recently used for regression  or classification analysis with high-dimensional compositional covariates \citep{bates2019log,monti2021robust,gordon2022learning}. 
	The general form of these models is given by
	\begin{equation}
		\label{eq:logcontrastreg}
		\mathbf{y} = \ln(\mathbf{X}) \cdot \boldsymbol{\beta} + \boldsymbol{\varepsilon},
	\end{equation}
	where $\mathbf{y} = (y_{1},\dots,y_{n})^\top$ is a univariate response variable, $\mathbf{X}=(x_{ij})_{1\leq i \leq n, 1\leq j \leq D}$ is a matrix of the compositional predictor, $\boldsymbol{\beta}$ represents the regression coefficients with $\sum_{j=1}^{D}\beta_{j} = 0$, and $\boldsymbol{\varepsilon}$ stands for the ordinary random error term. It is common practice to interpret the regression coefficients directly in terms of the original parts like in a standard multiple regression model. Nevertheless, from a compositional perspective, just one logcontrast is estimated. 
	
	The model in Eq. (\ref{eq:logcontrastreg}) can, however, be immediately generalized to a setting where more orthonormal logcontrasts are estimated simultaneously. In order to reduce the dimension, it would be interesting to rank them according to decreasing relevance to explain or predict the response variable. Moreover, simplifying the interpretation of these logcontrasts would be beneficial. This is achieved by the novel PLS-PB method introduced in this work as showed in the following.
	
	\subsection{Log-ratio Representations of Compositional Data}
	As said, CoDa are commonly expressed in the form of log-ratios of parts (logcontrasts) for statistical analysis. 
	Thus, clr coefficients \citep{aitchison1986statistical} are often used. For a $D$-part composition $\mathbf{x}=(x_1,\ldots,x_D)^\top$, its representation in the clr coefficients is defined as
	\begin{equation}
		\label{eq:clr}
		\textrm{clr}(\mathbf{x}) = \left( \textrm{ln}\frac{x_{1}}{g(\mathbf{x})}, \hspace{2pt} \textrm{ln}\frac{x_{2}}{g(\mathbf{x})}, \dots, \hspace{2pt} \textrm{ln}\frac{x_{D}}{g(\mathbf{x})} \right),
	\end{equation}
	where $g(\mathbf{x})$ denotes the geometric mean of the parts of the composition $\mathbf{x}$. Clr coefficients and logcontrasts are closely linked together as it holds that \citep{martin2018advances}
	\begin{equation}
		\label{eq:clrlog}
		\sum_{i=1}^{D}a_{i}\textrm{clr}_{i}(\mathbf{x}) = \sum_{i=1}^{D}a_{i}\ln x_{i},  \quad \sum^{D}_{i=1} a_{i} = 0,
	\end{equation}
	\noindent
	where $\textrm{clr}_{i}(\mathbf{x}) = \textrm{ln}\frac{x_{i}}{g(\mathbf{x})}$. The clr coefficients impose a zero sum constraint and thus lead to a singular covariance matrix, which is undesirable for some statistical methods including LS regression or LDA/QDA. However, their construction and interpretation is appealing for some others including PLS regression. Note that, by taking Eq. (\ref{eq:clrlog}) into account, the regression model in Eq. (\ref{eq:logcontrastreg}) could have been developed directly in clr coefficients, but we will do it separately later. The reason is methodological: while in Eq.  (\ref{eq:logcontrastreg}) the zero-sum constraint is additionally imposed which needs to accommodate estimators of regression coefficients used, like in \cite{monti2021robust}, with applying clr coefficients first it is automatically incorporated to proceed with standard estimators, but care needs to be taken for the interpretation of this log-ratio coordinate representation.
	
	An alternative way to map CoDa from their original sample space into the real space is through orthonormal log-ratio (olr) coordinates, also known as isometric log-ratio (ilr) coordinates, which are derived from the Euclidean vector space structure of the Aitchison geometry and overcome the singularity issue, hence being more generally applicable in statistical analysis \citep{egozcue2003isometric,filzmoser18,martin19}.
	
	Note that balances are one concrete instance of olr coordinates that are constructed by means of a sequential binary partition (SBP) of the parts of a composition \citep{egozcue05}. This procedure sequentially splits parts into two non-overlapping groups. Thus, at the $k$th partition, $k = 1,\dots,D-1$, the balance $b_{k}$ between two subgroups is given by
	$$
	b_{k} = \sqrt{\frac{r_{k}s_{k}}{r_{k}+s_{k}}}\text{ln}\frac{\sqrt[\leftroot{-1}\uproot{2}r_{k}]{x_{n_{1}}\cdot \dots \cdot x_{n_{r_{k}}}}}{\sqrt[\leftroot{-1}\uproot{2}s_{k}]{x_{d_{1}}\cdot \dots \cdot x_{d_{s_{k}}}}},
	$$
	\noindent
	where $r_k$ denotes the number of parts in the first group (numerator of the log-ratio) and $s_k$ denotes the number of parts in the second group (in the denominator of the log-ratio), with $n_{1},\dots,n_{r_{k}}$ and $d_{1},\dots,d_{s_{k}}$ being the indices of the parts of the first and second group, respectively. From a $D$-part composition, the number of balances derived is $D-1$ which corresponds to the actual dimensionality of the  composition. The interpretation of balances is straightforward: they represent the relative dominance of one group of parts with respect to the other group. They are orthonormal logcontrasts by construction, which means that the respective vectors of logcontrast coefficients are re-scaled to have unit norm and are mutually orthogonal:
	
	$$
	b_{k}=\sum_{i=1}^{D}a_{ki}\ln x_{i},\;\text{with } a_{ki}=
	\begin{cases}
		\sqrt{\frac{s_i}{(r_i + s_i)r_i}}     & \text{if } i \in \{ n_1,\dots,n_{r_{k}} \}  \\
		-\sqrt{\frac{r_i}{(r_i + s_i)s_i}}    & \text{if } i \in \{ d_1,\dots,d_{s_{k}} \} \\
		\quad 0 & \text{otherwise}.
	\end{cases}
	$$

	Recently, the use of balances has been questioned by some \citep{greenacre19,greenacre21,hron2021analysing}. Nonetheless, their ability to represent the original information in terms of contrasts or comparisons between two groups of parts remains appealing when compared to using general logcontrasts, particularly so in high-dimensional settings.
	
	However, as the number $D$ of parts of a composition increases it becomes more challenging to build a SBP and obtain a collection of balances that are interpretable. It is then desirable to construct just a few balances which capture the majority of the information. In an unsupervised learning settings, principal balances have been proposed for this aim \citep{pawlowsky2011principal}. In \cite{martin2018advances} PBs are formally defined as follows.
	
	\begin{definition}
		Given a composition $\boldsymbol{x} = (x_1,x_2,\dots,x_D)^\top$, principal balances are logcontrasts $\sum_{i=1}^{D} a_{ki} \cdot \ln x_{i}$, $k = 1,\dots, D-1$, such that $\boldsymbol{a}_{k} = (a_{k1},\dots,a_{kD})^\top$ are constant vectors which maximize the variances $\text{var}\left[\sum_{i=1}^{D}a_{ki} \cdot \ln x_{i}\right]$ and:
		\begin{itemize}
			\setlength\itemsep{1em}
			\item for $k = 1,\dots, D-1$ the coefficients $a_{ki}$ take one of the three values $(-c_1,0,c_2)$, $c_1$ and $c_2$ being some strictly positive numbers,
			\item for $k = 1,\dots, D-1$, it holds that $\sum_{i=1}^{D}a_{ki} = 0 \quad \text{and} \quad \sum_{i=1}^{D}a_{ki}^2 = 1$,
			\item for $k = 2,3,\dots, D-1$, $\boldsymbol{a}_{k}$ is orthogonal to the previous $\boldsymbol{a}_{k-1},\boldsymbol{a}_{k-2},\dots,\boldsymbol{a}_{1}$, that is $\sum_{i=1}^{D} a_{ki}\cdot a_{(k-l)i} = 0, \quad l = 1,2,\dots,k-1$.
		\end{itemize}
	\end{definition} 
	
	PBs facilitate dimension reduction using PCA in the clr space, but the interpretation of the resulting principal components is simplified through their expression in terms of balances. Accordingly, the first PB maximizes the sample variance and each subsequent balance then maximizes the remaining variance in the data, while satisfying the orthonormality constraint. Up to $D-1$ PBs can be derived, although in practice much fewer are typically needed to capture the main modes of variability. In the following subsection, we embed such dimension reduction by PB coordinates into a regression setting using a PLS formulation.
	
	\subsection{PLS Regression and Classification}
	
	PLS is a multivariate method which is used to model a linear relationship between a response variable and a set of (non necessarily compositional) explanatory variables. The linear relationship is however not modeled directly, but via the construction of latent variables (PLS components). Values of new latent variables are called scores, and  coefficients that determine the influence of each variable on the score are called loadings \citep{varmuza2009introduction}. 
	
	More precisely, PLS regression aims to estimate the regression parameter vector $\mathbf{b}=(b_1,\dots,b_D)$ in the linear 
	regression model 
	\begin{equation}
		\label{eq:regmodel}
		\mathbf{y} = \mathbf{X}\mathbf{b} + \mathbf{e},
	\end{equation}
	\noindent
	where $\mathbf{e}$ stands for a random error term. 
	Both the response variable and covariates are centered prior to the analysis, so no intercept is included in model \ref{eq:regmodel}. As we deal with CoDa, the matrix $\mathbf{X}$ in Eq. (\ref{eq:regmodel}) is expressed in the form of clr coefficients and this clr matrix is denoted $\textrm{clr}(\mathbf{X})$ in the following. 
	While the estimation of $\mathbf{b}$ in terms of the original variables (here clr coefficients as used for compositional PCA) is the final goal, the regression fit itself and prediction are performed on the PLS components, which are linear combinations of clr variables in the $n\times D$ matrix $\textrm{clr}(\mathbf{X})$. Because of Eq. (\ref{eq:clrlog}), these latent PLS components are just the logcontrasts we are searching for. 
	
	Namely, the matrix $\textrm{clr}(\mathbf{X})$ is decomposed as 
	$$
	\textrm{clr}(\mathbf{X}) = \mathbf{T}\mathbf{P}^\top + \mathbf{E}_{X},
	$$
	where $\mathbf{T}$ is a score matrix, $\mathbf{P}$ is a loading matrix, and $\mathbf{E}_{X}$ is an error matrix (\citealp{varmuza2009introduction}). Both matrices $\mathbf{T},\mathbf{P}$ have $k$ columns, $k \leq \textrm{min}(D,n)$, indicating the number of PLS components. 
	The goal of PLS is then to maximize the covariance between the scores (coordinates corresponding to the latent variables) and $\mathbf{y}$, under the constraint of uncorrelated scores (the most usual case) or orthogonal loading vectors (representing weights given to the original variables in the construction of the PLS components). Let $\mathbf{p}$ denote a (column) loading vector of matrix $\mathbf{P}$. Then it holds for a score vector $\mathbf{t}$ that $\mathbf{t} = \textrm{clr}(\mathbf{X})\mathbf{p}$, and the maximization problem can then be written as
	\begin{equation}
		\label{eq:max}
		\begin{aligned}
			\max_{\mathbf{p}} \quad & \textrm{cov}(\textrm{clr}(\mathbf{X})\mathbf{p},\mathbf{y}),
			\quad \textrm{subject to} \quad \lVert \textrm{clr}(\mathbf{X})\mathbf{p} \rVert = 1,
			\\
		\end{aligned}
	\end{equation}
	where $\mathbf{p}$ is considered to be a weighting vector. 
	The constraint of unit length ensures that the maximization problem is unique \citep{varmuza2009introduction}. Such maximization problem results in the first score vector $\mathbf{t}$. The subsequent score vectors are obtained in the same way, with the condition that they must be orthogonal to the previous ones. The score vector $\mathbf{t} =\textrm{clr}(\mathbf{X})\mathbf{p}$
	contains $n$ observations of a logcontrast because the sum of elements of each loading vector $\mathbf{p}$ equals to zero in the clr coefficient representation of the composition acting as covariate.
	The resulting logcontrasts can then be used in the regression model
	\begin{equation}
		\label{eq:plsmodel}
		\mathbf{y} = (\mathbf{T}\mathbf{P}^\top)\mathbf{b} + \mathbf{e}_T=\mathbf{T}\mathbf{v} + \mathbf{e}_T,
	\end{equation}
	with $\mathbf{v}=\mathbf{P}^\top\mathbf{b}$, for prediction purposes as well as to determine the number of logcontrasts that are sufficient for a good prediction of $\mathbf{y}$. There exist several algorithms to find a solution to this optimization problem. We resort to the well-known SIMPLS algorithm. 
	
	Similarly, PLS can be used for classification purposes. This method is then commonly called PLS Discriminant Analysis (PLS-DA). Here we will focus on binary response variables, typically using codes 0 for observations that do not belong to a certain group and 1 for those that do belong.
	
	Beyond representing a generalization of logcontrast models, a step further with the proposed formulation is to simplify it in the form of balances, and then rely on PBs instead of PLS loadings. Thus, we can investigate which groups of parts contribute in positive or negative sense to their values. In doing so, a small price is paid in terms of prediction ability of the resulting regression model, but
	a benefit in interpretability of logcontrasts as latent variables is generally obtained.

	%%%%%%%%%%%%%%%%%%%%%%%%%%%%%%%%
	\subsection{Algorithmic Implementation of PLS Principal Balances}   \label{PropAlg}
	PLS-PB can be straightforwardly constructed by adapting the  PCA-PB approach from \cite{martin2018advances}. Specifically, we take the constrained PCs algorithm introduced in that work as reference. This algorithm builds PBs based on loadings obtained from PCA. Our proposal is to do it based on the loadings from PLS.%}

The core of the modification is as follows: instead of maximizing explained variance in accordance with PCA, we maximize the covariance between the response variable and the new established balance (through the respective balance coefficients) while keeping orthonormality of the new coordinate system. Thus, balances and balance coefficients play the role of score vectors and loadings respectively. In the relationship $\mathbf{t} = \textrm{clr}(\mathbf{X})\mathbf{p}$, balance coefficients are in the place of vector $\mathbf{p}$. Accordingly, the first PLS loading vector is used to derive the first PB, and the other balances are then obtained by maximizing  the absolute value of the covariance of subsequently derived balances with the response variable. Algorithm \ref{algorithmPLSPB} summarizes this procedure for obtaining the PLS-PB.

%%% NEW VERSION
\begin{algorithm}
	%\setstretch{1.1}
	\caption{CONSTRUCTION OF PLS-PB}
	\label{algorithmPLSPB}
	% Initiation section
	\smallskip
	\textbf{Initiation}: center response variable $\boldsymbol{y}$, compute clr coefficients of composition in $\boldsymbol{X}$ and center them
	
	\medskip
	% Construction of PBs section
	\textbf{PB:}
	\begin{enumerate}
		\item First PB: $\boldsymbol{pb}_{1}$ (based on first PLS loading vector $\boldsymbol{p}_{1}$): 
		\begin{enumerate}
			\item[i.] PLS regression of centered $\boldsymbol{y}$ on centered clr$(\boldsymbol{X})$
			\item[ii.] First loading vector:  $\boldsymbol{p}_1$
			\item[iii.] Signs of values in $\boldsymbol{p}_1$: $\boldsymbol{s}_{sign}=\textrm{sign}(\boldsymbol{p}_1)$
			\item[iv.] Using $\boldsymbol{p}_1$, derive $D-1$ candidate sign of balances  $\boldsymbol{s}_{1},\dots,\boldsymbol{s}_{D-1}$ with codes $\{-1,0,+1\}$:
			\begin{itemize}
				\item for $i = 1,\dots,D$:
				$$
				s_{i1}=
				\begin{cases}
					+1     & \text{if $p_{i} = \max_{ \{ 1 \leq i \leq D \} } \boldsymbol{p}_1$} \\
					-1     & \text{if $p_{i} =  \max_{ \{ 1 \leq i \leq D \}} (- \boldsymbol{p}_1)$} \\
					\;\; 0      & \text{otherwise.}
				\end{cases}
				$$
				\item $\boldsymbol{s}_{j}$, $j = 2,\dots,D-1$: copy codes in $\boldsymbol{s}_{j-1}$, add $+1$ or $-1$ using $\boldsymbol{s}_{sign}$ and the remaining components of $\boldsymbol{p}_{1}$ (excluding the minimum and maximum values chosen in the previous step). For $i = 1,\dots,D$ :
				$$
				s_{ij}=
				\begin{cases}
					s_{i(j-1)}     & \text{if } s_{i(j-1)}\neq 0   \\
					sign(\boldsymbol{p}_{1i})     & \text{if } \lvert p_{1i}\rvert = \max_{ \{ k: s_{k(j-1)}=0 \} } \{ \lvert p_{1k} \rvert\}   \\
					\quad 0      & \text{otherwise.}
				\end{cases}
				$$
				\item result: matrix of signs $\boldsymbol{S}_{D \times (D-1)}$; sign of balances in columns
			\end{itemize}
			\item[v.] Create $\boldsymbol{B}_{D \times (D-1)}=(\boldsymbol{b}_{1},\dots,\boldsymbol{b}_{(D-1)})$ matrix of balance coefficients: for $i^{th}$ row and $j^{th}$ column, $i = 1,\dots,D$, $j = 1,\dots,D-1$
			$$
			b_{ij}=
			\begin{cases}
				\sqrt{\frac{s_j}{(r_j + s_j)r_j}}     & \text{if $s_{ij} = 1$,} \\
				-\sqrt{\frac{r_j}{(r_j + s_j)s_j}}    & \text{if $s_{ij} = -1$,} \\
				\quad 0 & \text{otherwise.}
			\end{cases}
			$$
			where $r_j$ is number of $+1$ values in column $j$ and $s_j$ is number of $-1$ values in columns $j$ \\
		\end{enumerate}
	\end{enumerate}
\end{algorithm}

\begin{algorithm}
	\ContinuedFloat
	\caption{CONSTRUCTION OF PLS-PB (continued)}
	\begin{itemize}
		\item[]\begin{enumerate}
			\item[vi.] First PB: 
			$\boldsymbol{pb}_{1} = \max_{ \{ 1\leq j \leq (D-1) \} }  \lvert \textrm{cov}(\textrm{ln}(\boldsymbol{X})\boldsymbol{b}_{j},\boldsymbol{y})\rvert$
		\end{enumerate}
	\end{itemize}
	
	\begin{enumerate}
		\item[2.] Derive the other balances based on $\boldsymbol{pb}_{1}$:
		\begin{enumerate}
			\item[i.] If $\boldsymbol{pb}_{1}$ contains $0$ value(s): repeat the procedure for $\boldsymbol{pb}_{1}$ (steps 1.[i.-vi.]) using only variables assigned with $0$ in $\boldsymbol{pb}_{1}$
			\item[ii.] Further partition: 
			\begin{itemize}
				\item Down the numerator: repeat steps 1.[i.-vi.] using variables in the numerator of $\boldsymbol{pb}_{1}$
				\item Down the denominator: repeat steps 1.[i.-vi.] using variables in the denominator of $\boldsymbol{pb}_{1}$
			\end{itemize}
		\end{enumerate}
	\end{enumerate}
	
	\medskip
	\noindent
	\textbf{Final step:}
	Sort PB: $\boldsymbol{pb}_{1},\boldsymbol{pb}_{2},\dots,\boldsymbol{pb}_{(D-1)}$ such that \\
	$\lvert \textrm{cov}(\boldsymbol{pb}_{(1)},\boldsymbol{y})\rvert > \lvert \textrm{cov}(\boldsymbol{pb}_{(2)},\boldsymbol{y})\rvert > \dots > \lvert \textrm{cov}(\boldsymbol{pb}_{(D-1)},\boldsymbol{y})\rvert$

\end{algorithm}

%%%%%%%%%%%%%%%%%%%%%%%%%%%%%%%%%%%%%%%%%%%%%%%%%%%%%%%%%%%%%%%%%%%%%%%%
\section{Numerical Assessment}    \label{sec:simulations}
%%%%%%%%%%%%%%%%%%%%%%%%%%%%%%%%%%%%%%%%%%%%%%%%%%%%%%%%%%%%%%%%%%%%%%%%
In this section, we first introduce some examples to demonstrate that PLS-PBs help to arrive at a simplified structure of PLS loadings. In line with an application to metabolomics in Section \ref{RealDat}, here we refer to (bio)markers (explanatory variables), i.e.\ biological measurements or signals most associated to some health or biological outcome/status of interest (response variable) \citep{stefelova2021weighted}. We therefore focus on the identification of meaningful biomarkers as the main purpose of the data analysis. Subsequently, we set up a simulation study to formally compare the predictive performance of the proposed PLS-PB with the original PCA-PB in \cite{martin2018advances}.

\subsection{Artificial Settings for Comparison: PLS-PB Against PLS Loadings} \label{Motivation}
\begin{figure}[t]
	\centering
	\subfigure[Original setting]{
		\label{subfig:lab1}
		\includegraphics[width=0.45\textwidth]{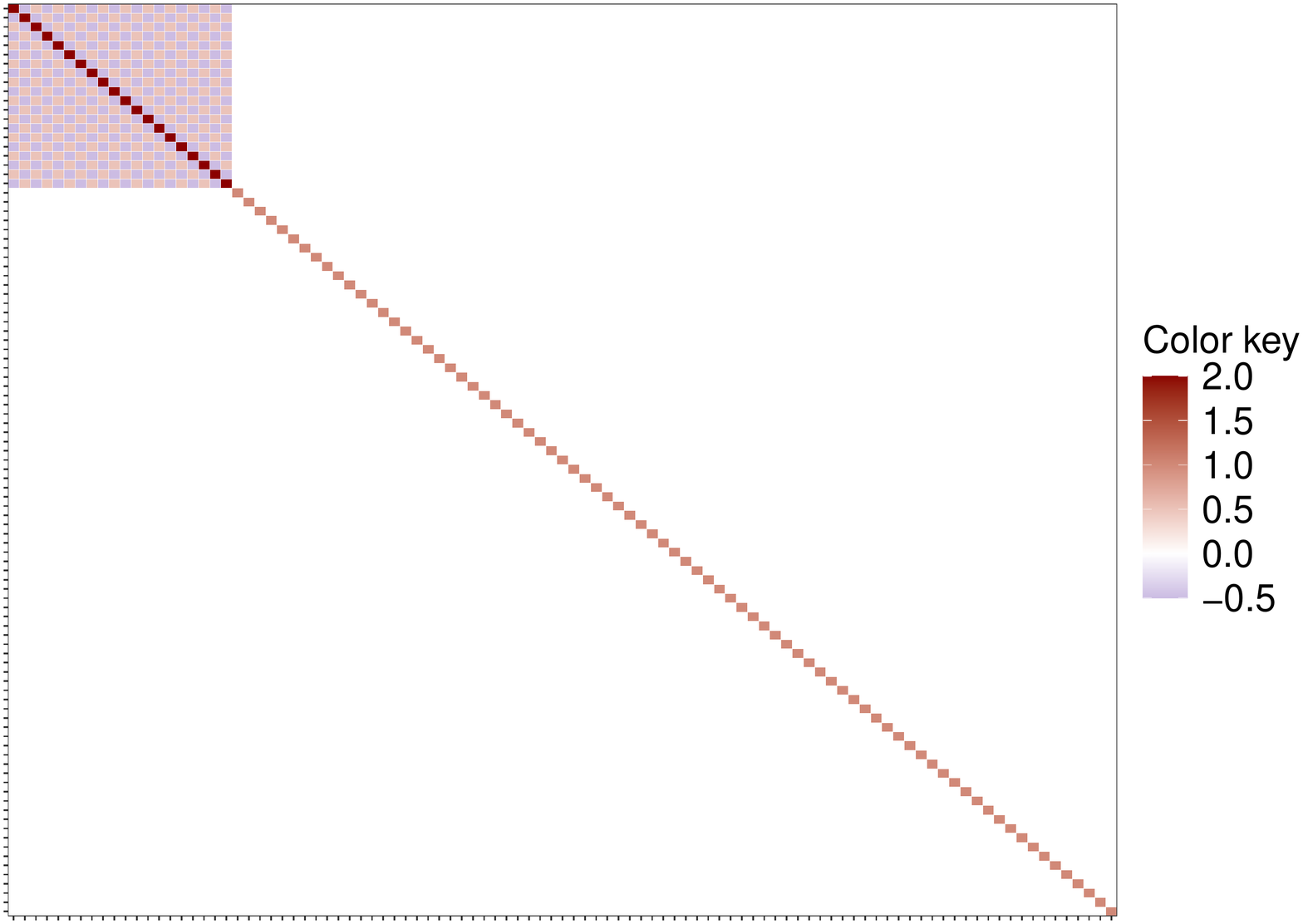}
	}
	\subfigure[Same-sized blocks]{
		\label{subfig:lab2}
		\includegraphics[width=0.45\textwidth]{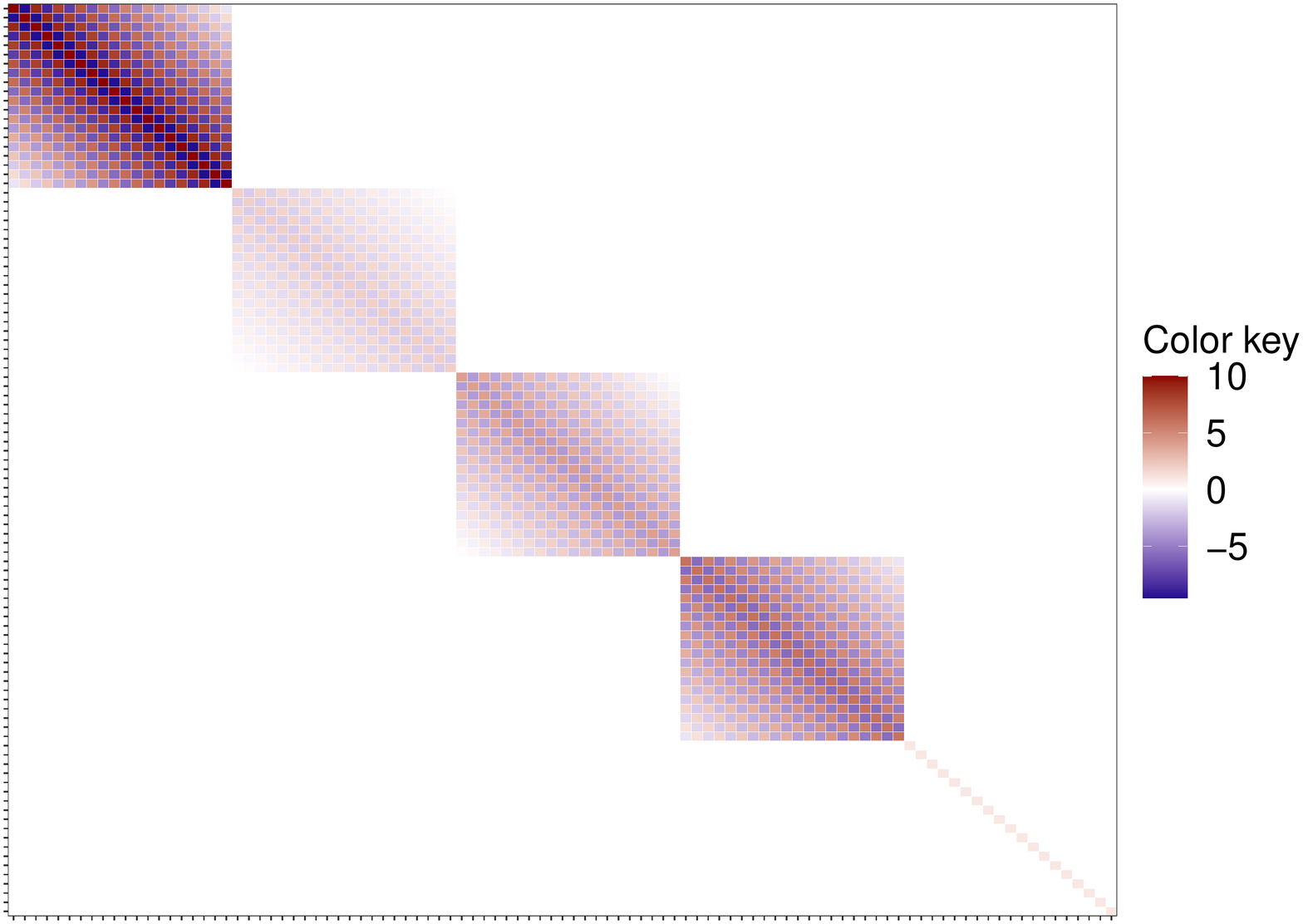}
	}
	\subfigure[Different-sized blocks]{
		\label{subfig:lab3}
		\includegraphics[width=0.45\textwidth]{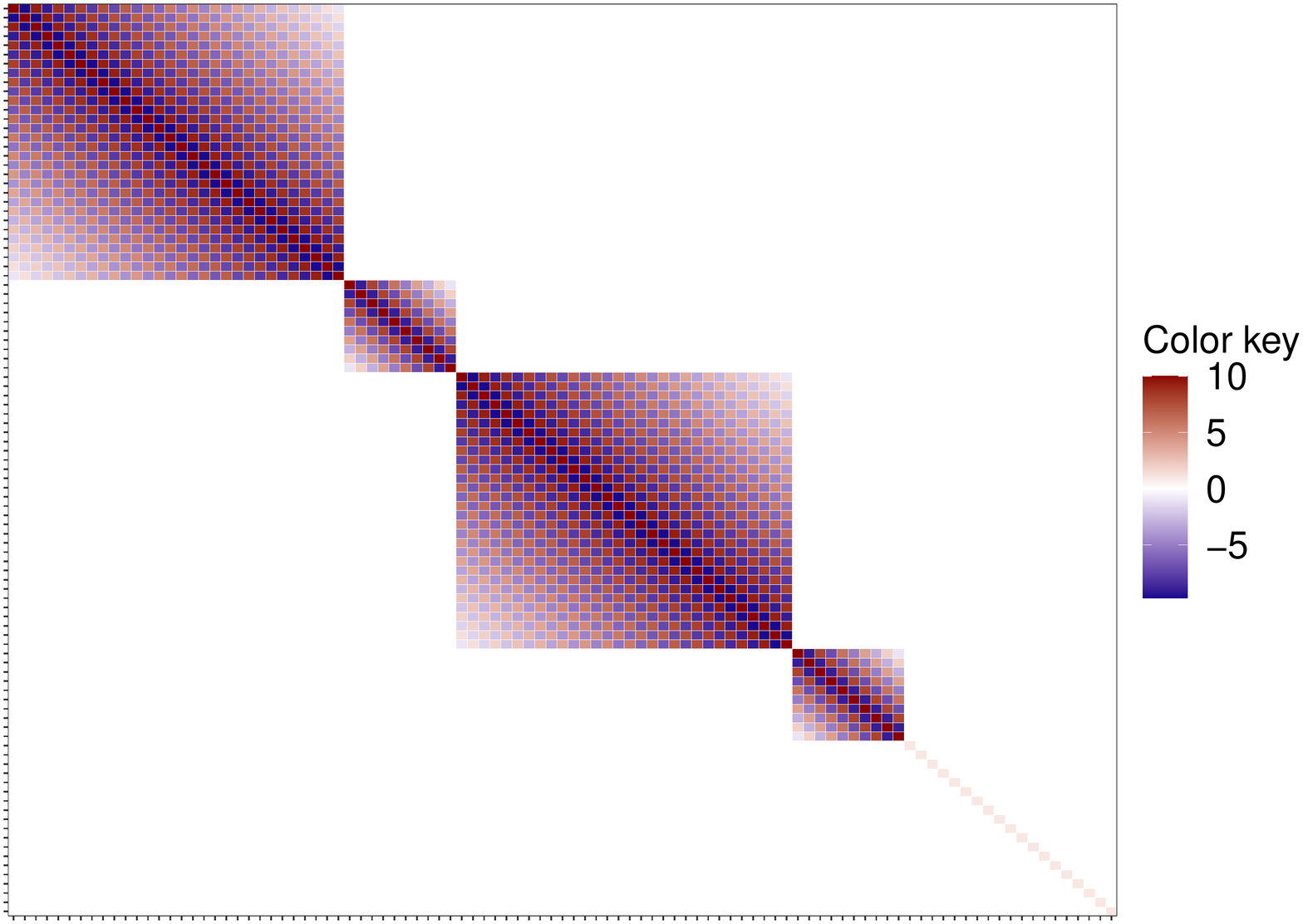}
	}
	\caption{Visual representation of covariance matrices used in the examples. Colors represent the covariance between pairs of pivot coordinates. Figure 1(a) shows a single block of $20$ meaningful markers, Figure 1(b) $4$ same-sized blocks of $20$ markers each and Figure 1(c) $4$ blocks of markers containing either $30$ or $10$ markers.}
	\label{fig:CovMat}
\end{figure}

To assess the behavior of PLS-PB across various settings, we consider three artificial cases inspired by the study in \cite{stefelova2021weighted}. The first case contains just one block of markers amongst a given collection of signals defining the explanatory composition. In the other two examples, we extend such setting to include several groups of markers. In all cases, we consider $n=250$ samples and $D=100$ signals in the composition. The number of PLS-PB is then $99$ ($D-1$). Following on \citep{stefelova2021weighted}, compositions were simulated using so-called pivot coordinates, an instance of olr coordinates \citep{FiHr2011} (see 
Appendix \ref{appA} for more details). 
The resulting covariance matrices are visualized in Figure \ref{fig:CovMat}.

Figure \ref{subfig:lab1} displays the covariance matrix in the first case. The generated compositions then contain one block of $20$ markers and the remaining signals are regarded unimportant.
For the other two exemplary cases, we adapt the covariance matrix defined in Eq. \eqref{simcov} of Appendix \ref{appA} to allow for multiple blocks of markers. In the second case, we consider four blocks of markers (each block consisting of $20$ markers) as visualized in Figure \ref{subfig:lab2}. Although the structure of the covariance matrix is the same as in Equation \eqref{simcov} (Appendix \ref{appA}), the elements in each block are generated following a decreasing sequence such that the elements further away from the diagonal have smaller values. This approach ensures that the scattering of pivot coordinates in each block differs. The effect of the first block of pivot coordinates is the strongest,  the second block should produce the ``weakest'' markers (in the sense of pivot coordinates; covariance between variables in the second block is the lowest). Because of the way pivot coordinates are built, it can be said that the relevance of markers (in terms of the original parts) decreases in each subsequent block. 

Finally, in the third case, we again consider four blocks of markers but this time of different size, as visualized in Figure \ref{subfig:lab3}. The first and third block consist of $30$ markers, while the second and fourth consist of $10$ markers each. The entries in the covariance matrix are in the same range in each block, with the diagonal elements being identical. Note that the second and third case not only contain 80 markers but also 20 irrelevant signals (i.e.\ the last 20 rows/columns in Figures \ref{subfig:lab2} and \ref{subfig:lab3}),  which can be considered as noise).

We are now ready to compare PLS-PB to PLS loadings. For this, it is important to note that the structure of PLS-PB must not necessarily reflect the structure of the corresponding PLS loadings. This is due to the fact that orthonormality is required when constructing PLS-PB, while this is not the case for PLS loadings. These examples provide a first insight into the performance of PLS-PB to correctly identify markers in a collection of signals. 

\begin{figure}[t]
	\centering
	\subfigure[PLS loadings]{
		\label{subfig:heat1}
		\includegraphics[width=0.45\textwidth]{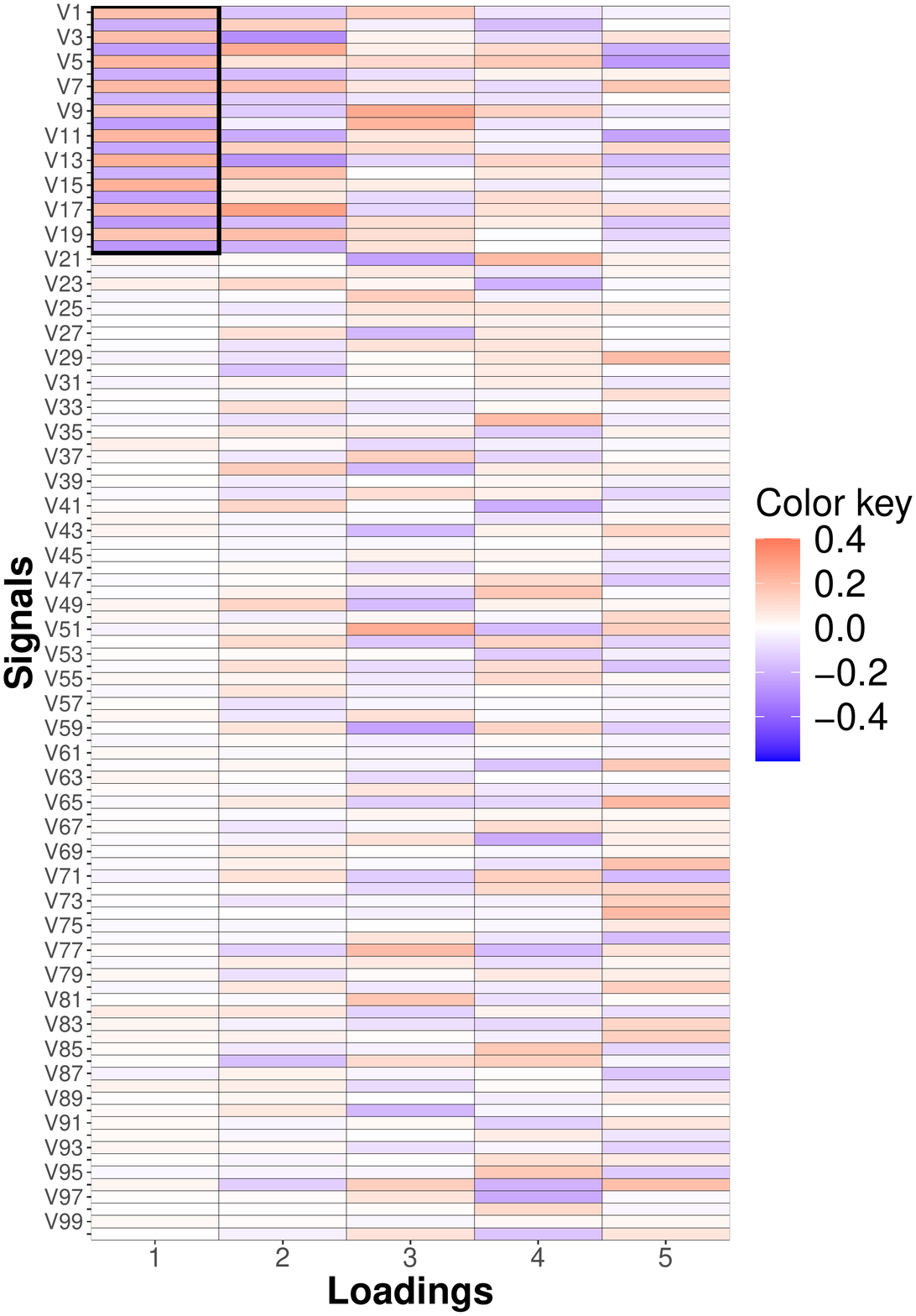}} 
	\subfigure[PLS-PB]{
		\label{subfig:heat2}
		\includegraphics[width=0.45\textwidth]{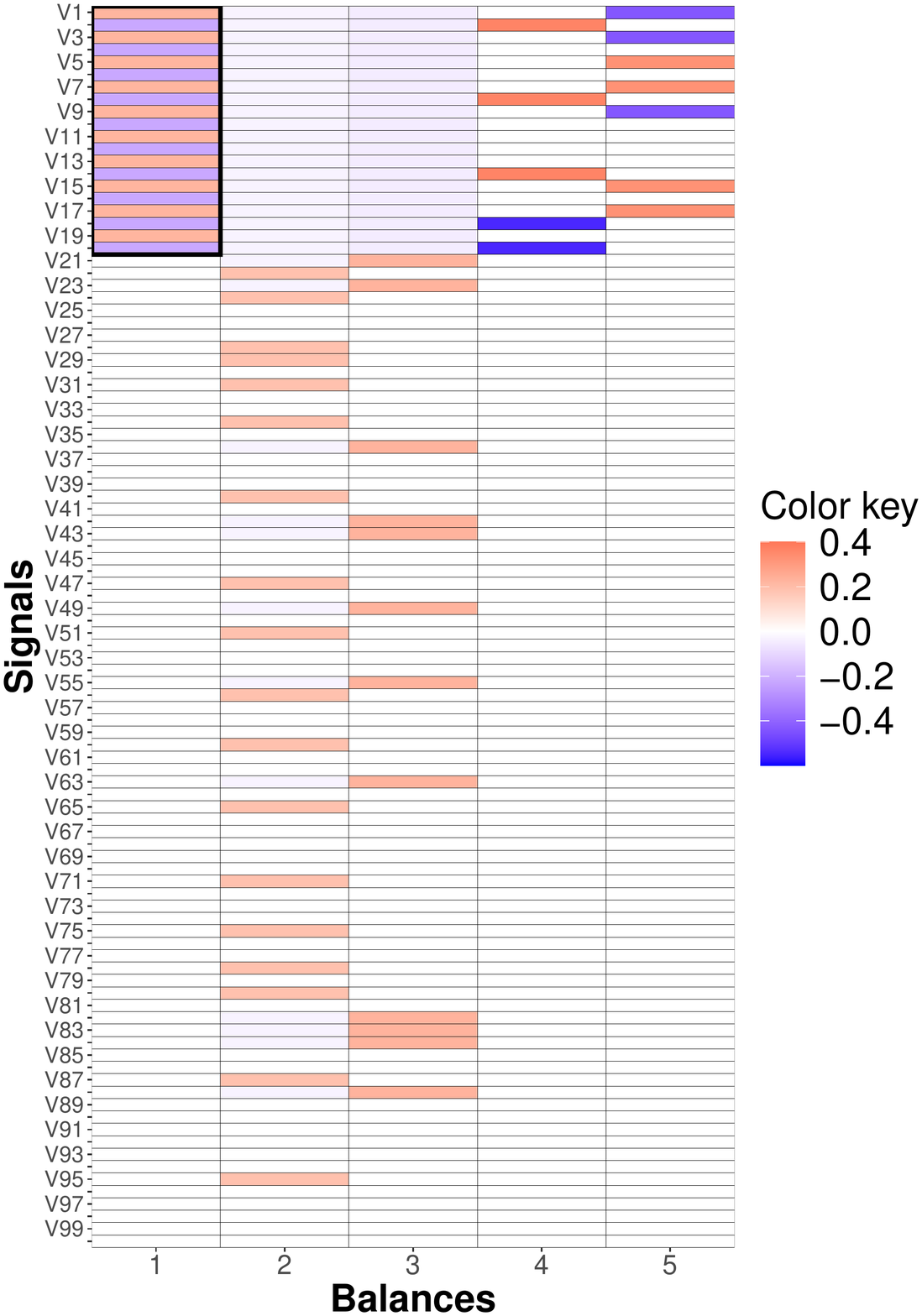}}
	\caption{Comparison of the first five PLS loadings (left) and PLS-PB (right) for the first example with one block of markers (case 1). Signal variables generated are arranged by rows. Colors represent values of loading vectors (left) and coefficients of PLS-PB (right). Signals in a numerator of a balance get a positive value, in a denominator a negative value. Signals not included in a balance get $0$. Block of markers (first $20$ signals) is highlighted into a black frame.}
	\label{fig:LoadPBsOrig}
\end{figure}

Figure \ref{fig:LoadPBsOrig} displays the comparison of PLS loadings and PLS-PB for the first case with one marker block. As we are typically interested in the first few PBs most strongly related to the response, we focus our discussion on the first five PLS loadings and PLS-PBs. These are displayed in the Figure \ref{subfig:heat1}, and Figure \ref{subfig:heat2} respectively. It can be clearly seen that the structure of PLS-PBs is much more parsimonious than using PLS loadings. The first PB (i.e.\ column 1 in Figure \ref{subfig:heat2}) captures the information contained in the first loading vector very precisely, highlighting all the markers in the data (i.e.\ first 20 colored rows in Figure \ref{subfig:heat2}). The fourth and fifth balances then capture several differences between markers in the block.

\noindent

\begin{figure}[t]
	\centering
	\subfigure[PLS loadings]{
		\label{subfig:heat1a}
		\includegraphics[width=0.45\textwidth]{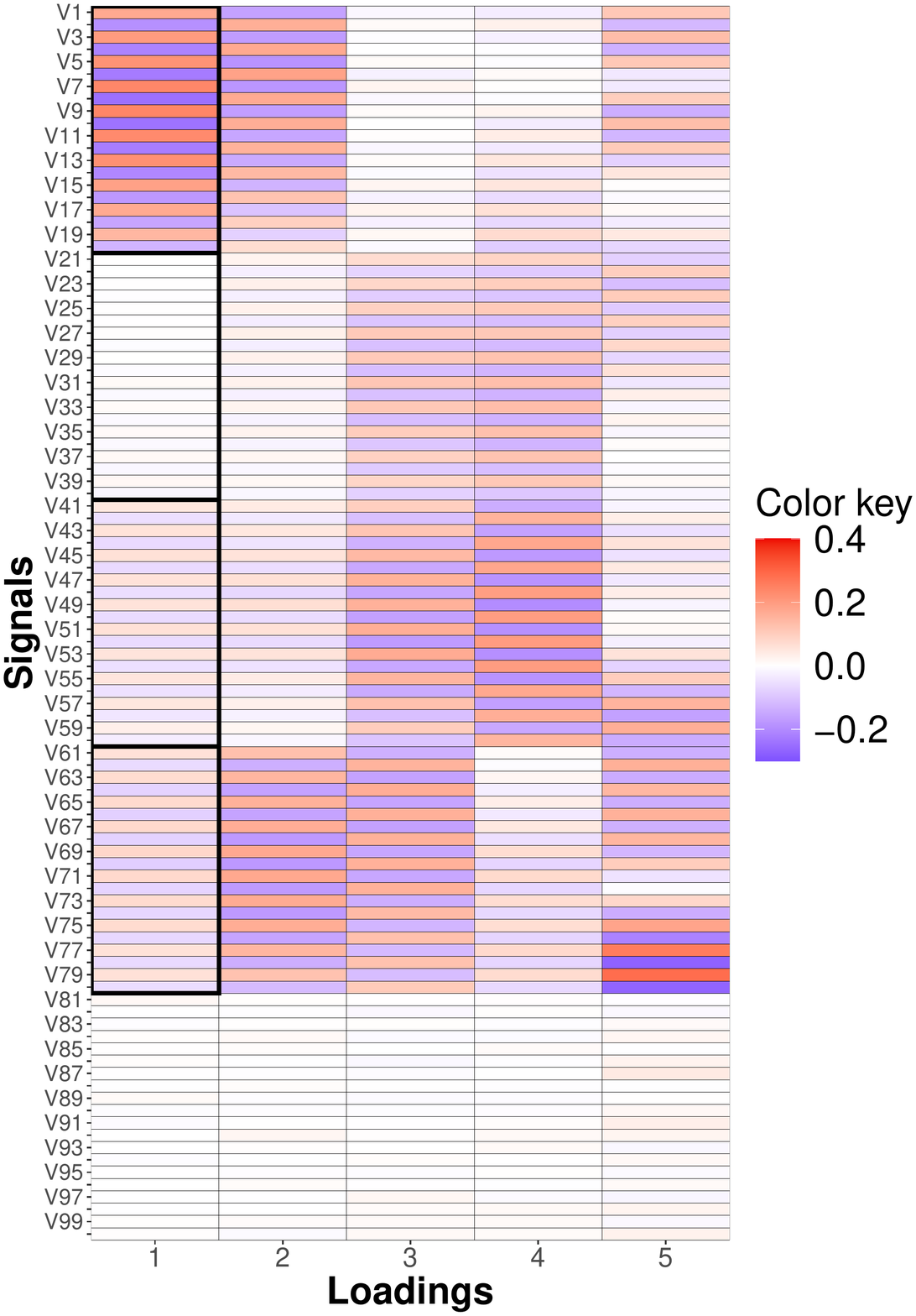}} 
	\subfigure[PLS-PB]{
		\label{subfig:heat2a}
		\includegraphics[width=0.45\textwidth]{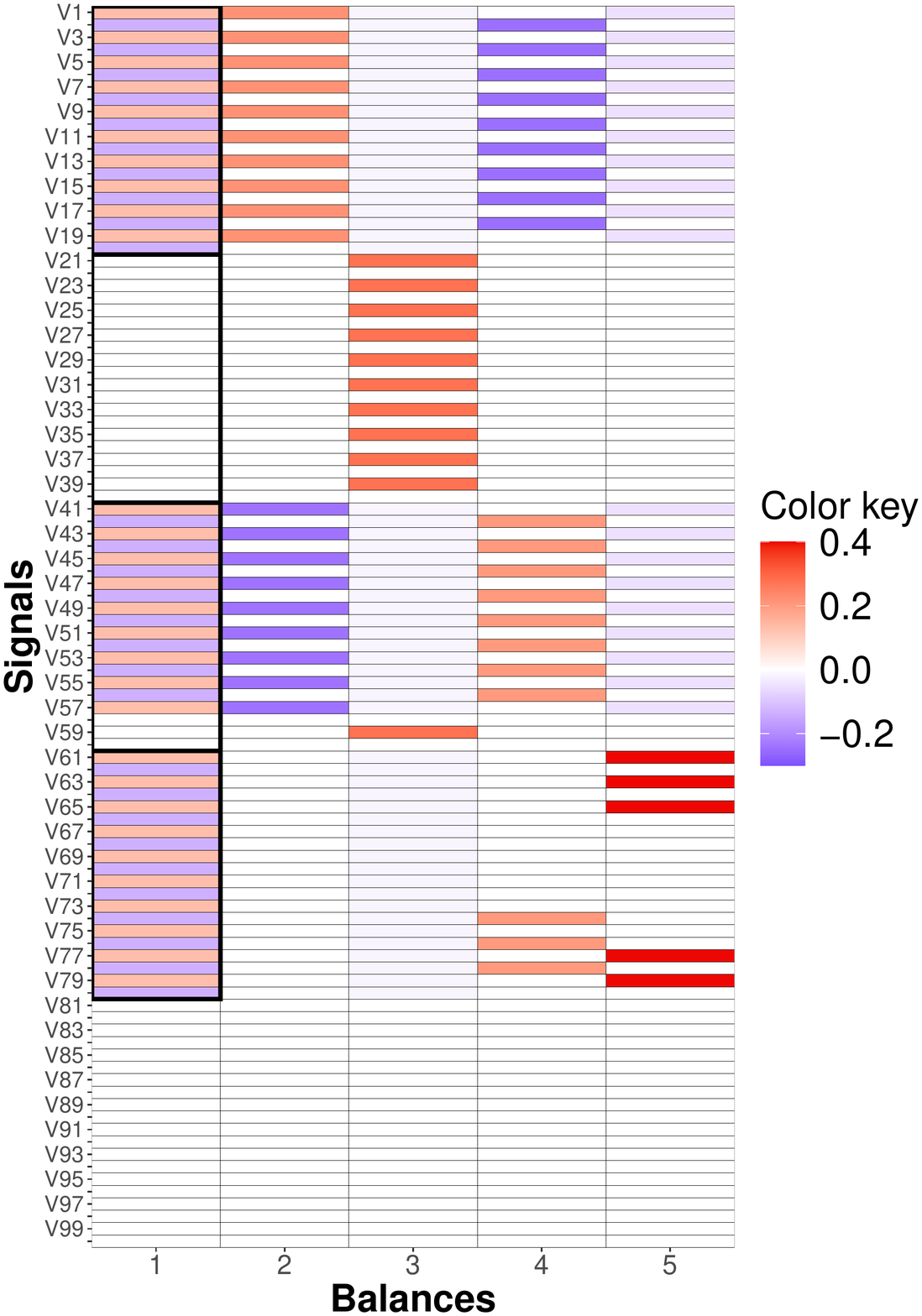}}
	\caption{Comparison of the first five PLS loadings (left) and PLS-PB (right) for the same-size blocks of markers setting (case 2). Signal variables generated are arranged by rows. Colors represent values of loading vectors (left) and coefficients of PLS-PB (right). Signals in a numerator of a balance get a positive value, in a denominator a negative value. Signals not included in a balance get $0$. Blocks of markers (signals V$1$-V$80$) are highlighted into a black frame.}
	\label{fig:LoadPBs1}
\end{figure}

Figure \ref{fig:LoadPBs1} displays the resulting PLS loadings and PLS-PB for case 2 with same-sized blocks of markers. The PLS-PBs in Figure \ref{subfig:heat2a} display a fairly neat structure, while the interpretation of the PLS loading in Figure \ref{subfig:heat1a} is less clear cut. The first PB reproduces the information in the first loading vector. Except for the second block of markers (signals from V$21$ to V$40$), the other three blocks are correctly identified. It is the third PB that highlights this block with the lowest covariances between variables. The second and fourth PB further stress the difference between the first and third block of markers, whereas the fifth PB captures some differences to the fourth block. Note that all PBs correctly exclude the signals corresponding to random noise  (i.e.\ the last $20$).

\noindent

\begin{figure}[h]
	\centering
	\subfigure[PLS loadings]{\includegraphics[width=0.45\textwidth]{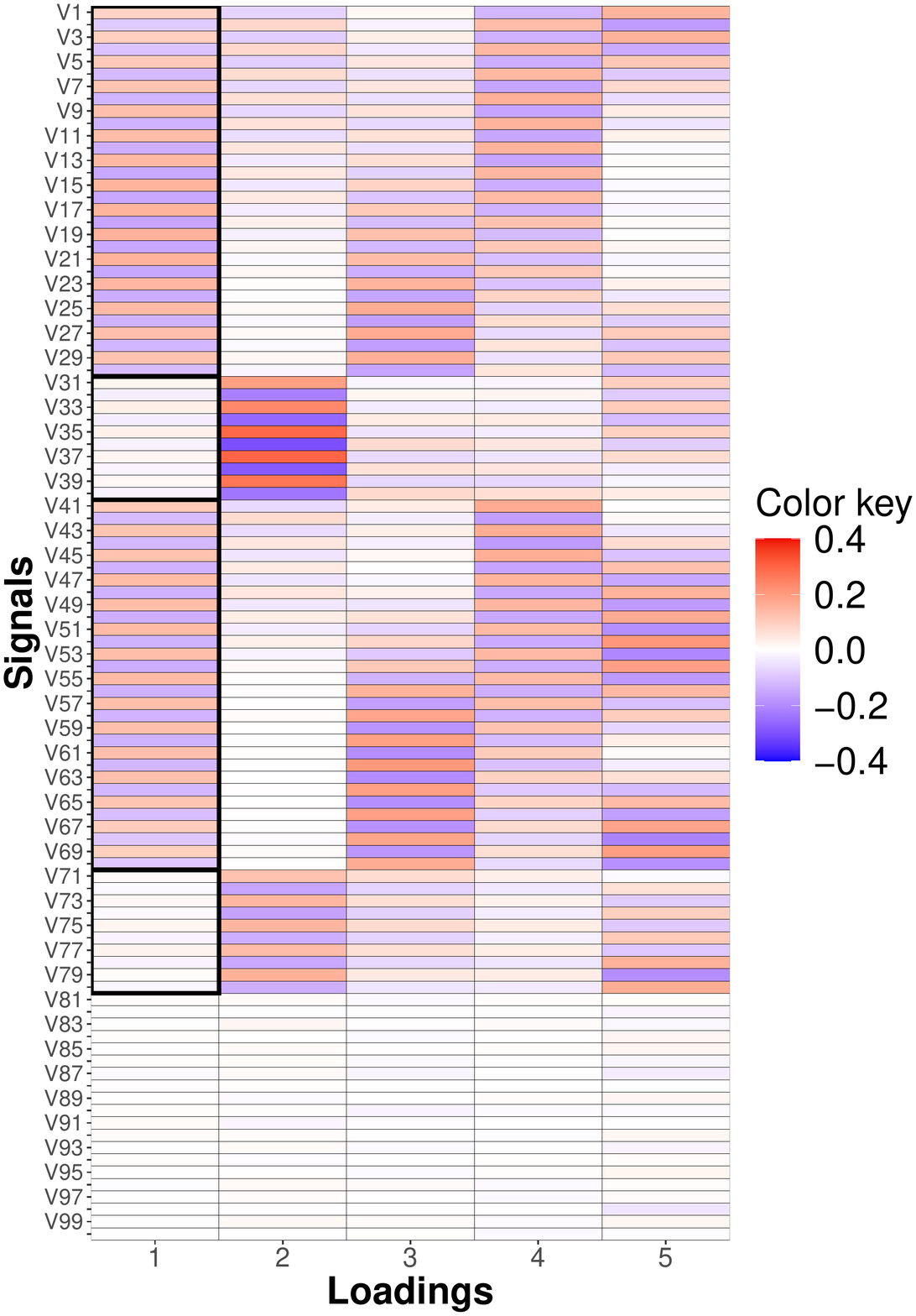}} 
	\subfigure[PLS-PB]{\includegraphics[width=0.45\textwidth]{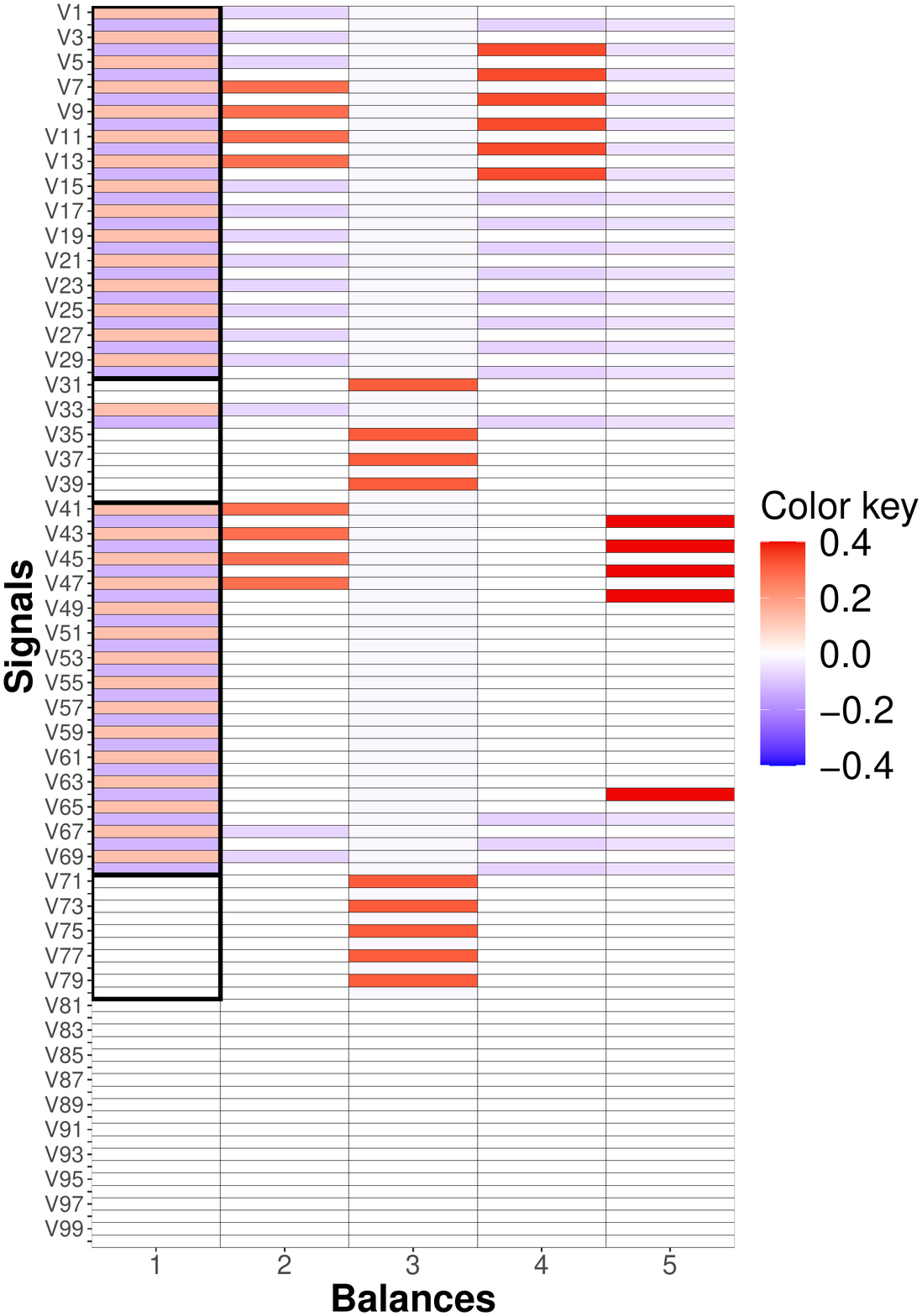}}
	\caption{Comparison of the first five PLS loadings (left) and PLS-PB (right) for the different-size blocks of markers setting (case 3). Signal variables generated are arranged by rows. Colors represent values of loading vectors (left) and coefficients of PLS-PB (right). Signals in a numerator of a balance get a positive value, in a denominator a negative value. Signals not included in a balance get $0$. Blocks of markers (signals V$1$-V$80$) are highlighted into a black frame.}
	\label{fig:LoadPBs2}
\end{figure}

Finally, Figure \ref{fig:LoadPBs2} displays the results for the case of varying block sizes (case 3). Similarly to the previous case, the largest blocks (i.e.\ the first and the third one; signals V$1$-V$30$ and V$41$-V$70$) are correctly highlighted by the first PB, while the second and the fourth (smaller) blocks are not included (signals V$31$-V$40$ and V$71$-V$80$ with the exception of signals number $33$ and $34$). However, markers from these blocks are identified by the third PB.

In summary, these examples illustrate the potential of PB as a convenient counterpart to PLS loadings, as they deliver a simplified structure and the orthonormality constraint enables to, for example, perform interpretable regression analysis \citep{hron2021analysing}. 

\subsection{Simulation-based Assessment}
\label{Sim}
We investigate the predictive ability of PLS-PB in comparison to PLS loadings, although it is important to note that this is not the main focus of the method proposed in this work. Moreover, a natural alternative to PLS-PB in a regression setting are PCA-based PBs as mentioned before. While PLS-PB correspond to PLS regression with a simplified loading structure, PCA-PB should follow the behavior of the well known principal component regression (PCR, \citealp{varmuza2009introduction}), where the number of explanatory variables in a regression model is reduced using PCA. Both PLS and PCA regression are popular tools, for example, in chemometrics and molecular biology applications to cope with high-dimensionality and/or multicollinearity issues. We therefore devise a simulation study to compare the prediction performance of PLS PB, PCA PB and PLS loadings. It is known  that PLS regression generally leads to better prediction performance than PCR when just a few latent components are involved. 

We consider three scenarios for the simulation study based on the three cases introduced in the previous section, that is: only one block of markers, several blocks of markers of the same size, and several blocks of markers of different sizes. The computed PBs (either PLS or PCA) are used to fit linear regression models like Eq. (\ref{eq:plsmodel}), including only one PB up to all possible PBs, that is, $99$. The \emph{root mean squared error of prediction}, 
\begin{equation*}
	\textrm{RMSEP} = \sqrt{\frac{1}{n} \sum_{i=1}^{n} (y_{i}-\hat{y_{i}})^2}, 
\end{equation*}
\noindent	
is used as prediction performance measure, where $y_{i}$ are the actual values and $\hat{y}_{i}$ the corresponding predicted values. This was estimated by $5$-fold cross-validation (CV) to provide a more realistic assessment, and it was evaluated over 100 simulation runs in each case. 

\begin{figure}[t]
	
	\centering
	\subfigure[One-block structure]{
		\label{subfig:lab1a}
		\includegraphics[width=0.47\textwidth]{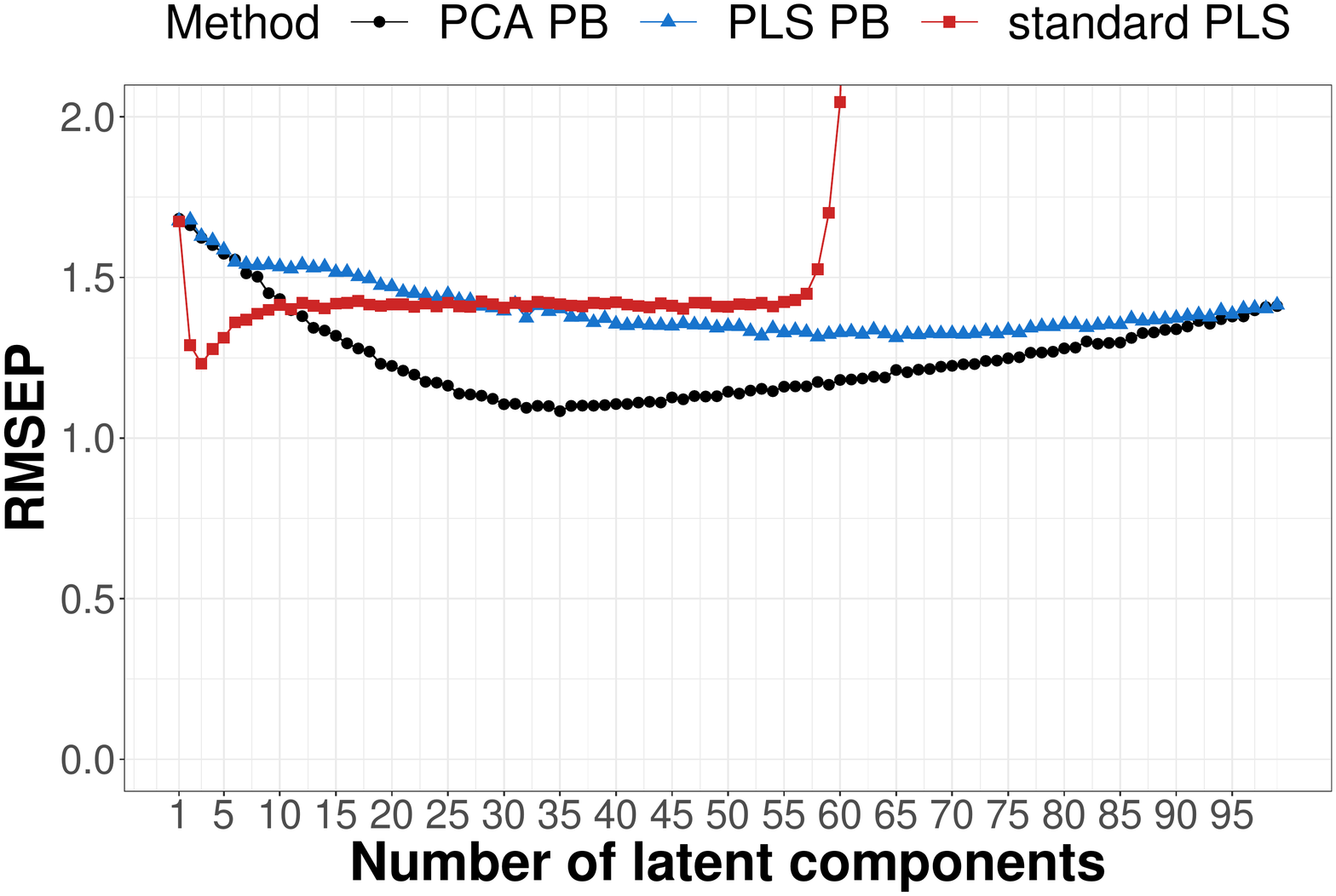}
	}
	\subfigure[Four same-sized blocks structure]{
		\label{subfig:lab2a}
		\includegraphics[width=0.47\textwidth]{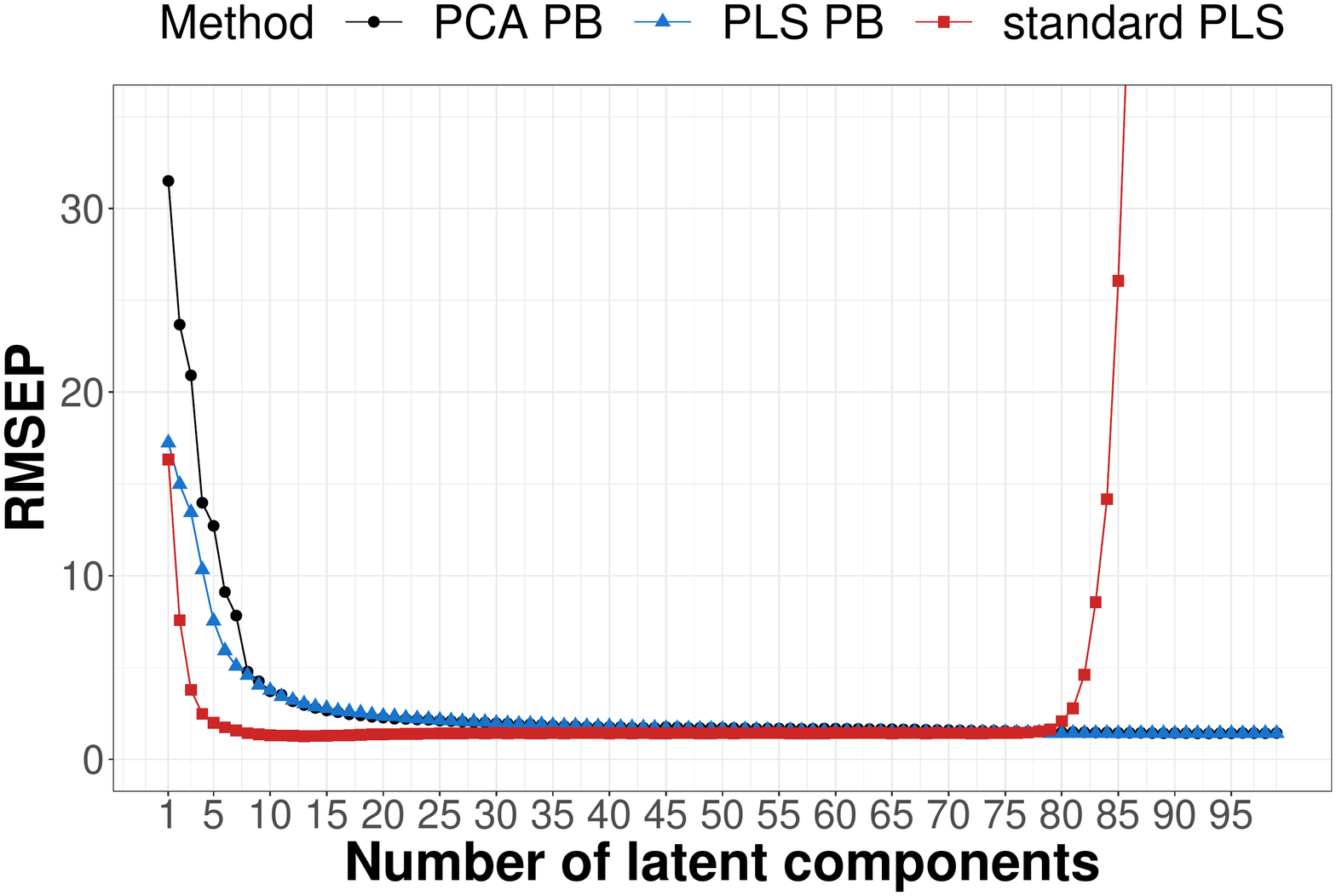}
	}
	\subfigure[Four different-sized blocks structure]{
		\label{subfig:lab3a}
		\includegraphics[width=0.47\textwidth]{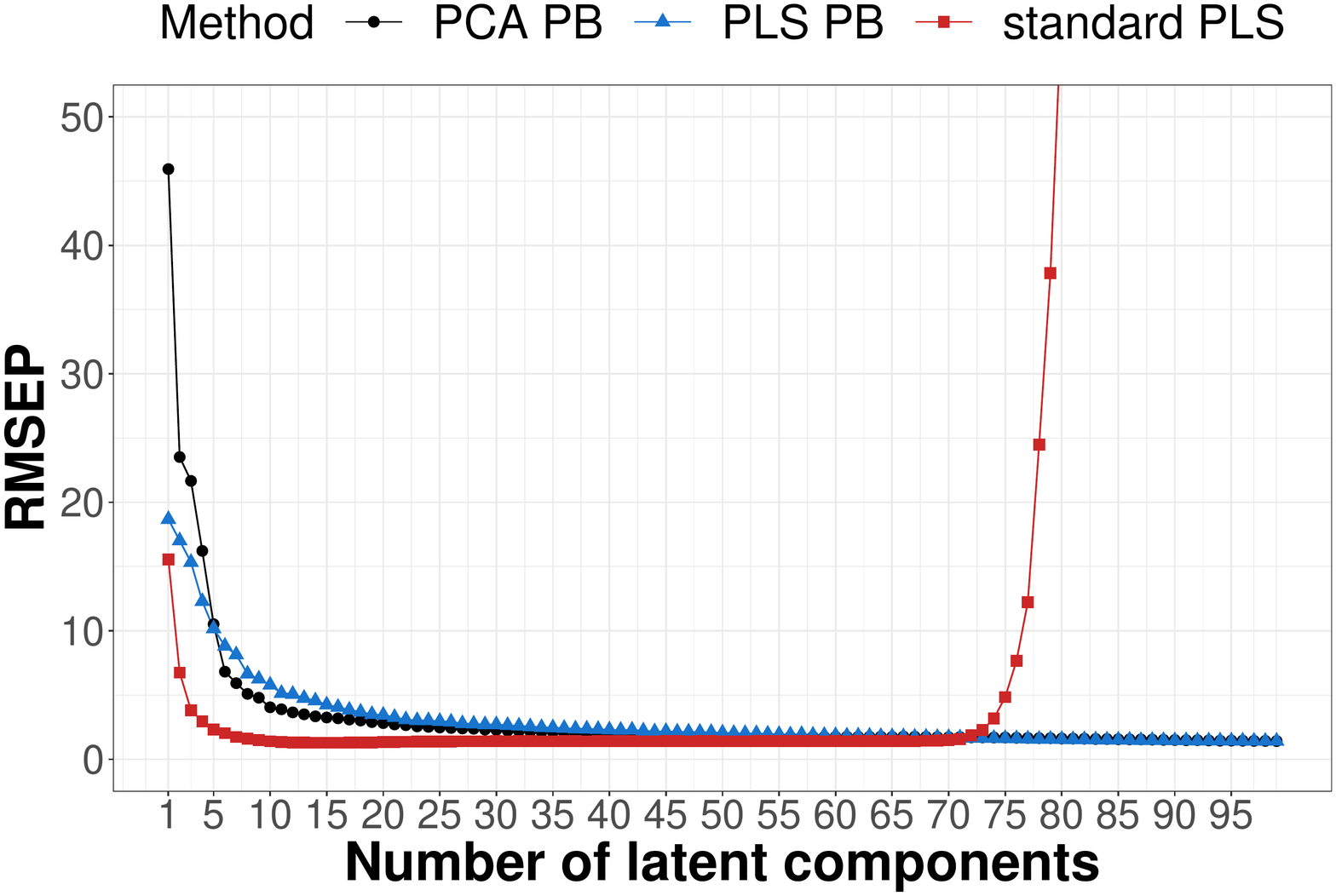}
	}
	
	\caption{Cross-validated root mean squared error of prediction using PB, PLS-PB (blue), PCA-PB (black), and ordinary PLS loadings (red) for each simulation scenario according to number of latent components used (either PB or ordinary loadings).}
	\label{fig:SimRuns}
\end{figure}

Figure \ref{fig:SimRuns} shows the results for all three simulation scenarios, comparing cross-validated RMSEP of PLS-PB against PCA-PB and PLS loadings. The number of PBs used (PLS loadings in case of standard PLS), ranging from one to $99$ ($D-1$) on the horizontal axis can be understood as an index of model complexity. Each point represents the average value of the RMSEP over the $100$ simulation runs.

First, we focus on the performance of PLS-PB against PCA-PB. When considering just a few PBs in the model, both perform very similarly in the one-block setting (Figure \ref{subfig:lab1a}), but PLS-PB considerably outperforms PCA-PB in the multiple marker settings (Figure \ref{subfig:lab2a} and Figure \ref{subfig:lab3a}). This is particularly relevant since in practice typically a small number of balances is preferred to facilitate interpretation of the results. Note that the value of the RMSEP  coincides for both PLS and PCA-PB when the maximum number of PBs is used because both systems of PBs are orthogonal rotations of each other, as it is the case with any other olr coordinate representation.

Looking now at differences between PLS-PB and standard PLS, it is not surprising that the latter performs best for the lowest numbers of latent components. Our aim is to construct interpretable PBs that explain most of the variation in the response and, in doing so, remain competitive in terms of predictive performance in relation to ordinary PLS. Hence, although the PB-based approach shows a slightly weaker prediction performance than standard PLS, it largely compensates this in terms of interpretability as demonstrated in Section \ref{Motivation}. Moreover, note that for models with a large number of latent components, the performance of standard PLS worsens dramatically, as reflected by the large values of the RMSEP. This might be caused by numerical instability resulting from applying the SIMPLS algorithm as it provides non-orthogonal components causing degeneration when increasing the number of components.

Lastly, we compare the proposed PLS-PB approach to the selbal algorithm introduced in \cite{rivera2018balances} to identify an optimal balance between parts in high-dimensional compositions for regression and classification problems in a microbiome analysis context. The latter cannot be included in Figure \ref{fig:SimRuns} as here we demonstrate prediction performance across all possible numbers of balances, and the selbal algorithm selects only one single balance. Nonetheless we can compare its performance to PLS-PB based on the first PB. The resulting mean values of the RMSEP for the three simulation scenarios are shown in Table \ref{tab:RMSEP}. While selbal exhibits better performance in terms of prediction, its ability to reflect the actual structure of markers in the data is notably poorer as discussed in the following.
\begin{table}[]
	\caption{\label{tab:RMSEP} Comparison of mean RMSEP of the first PCA PB and PLS PB with mean RMSEP of selbal balance.}
	\centering
	\begin{tabular}{|l|c|c|c|}
		\hline
		\multicolumn{1}{|c|}{Example }& PCA-PB & PLS-PB & selbal  \\ \hline
		1. Original setting  & 1.683 & 1.676 & 1.574 \\
		2. Same-sized blocks  & 31.507 & 17.243 & 10.671 \\
		3. Different-sized blocks  & 45.933 & 18.680 & 12.189 \\ \hline
	\end{tabular}
\end{table}

\begin{figure}[t]
	
	\centering
	\subfigure[One-block structure]{
		\label{subfig:lab1b}
		\includegraphics[width=0.3\textwidth]{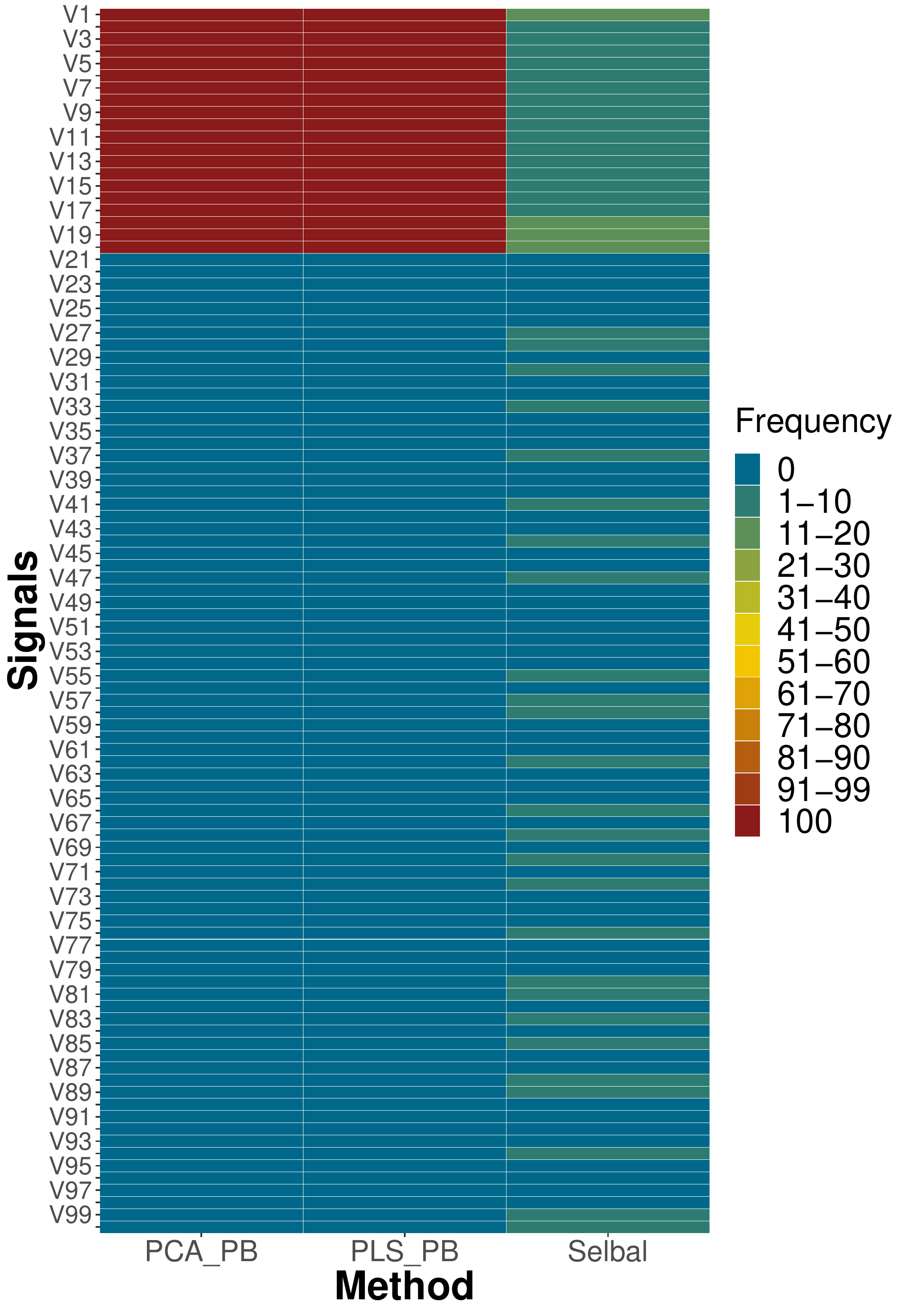}
	}
	\subfigure[Four same-sized blocks structure]{
		\label{subfig:lab2b}
		\includegraphics[width=0.3\textwidth]{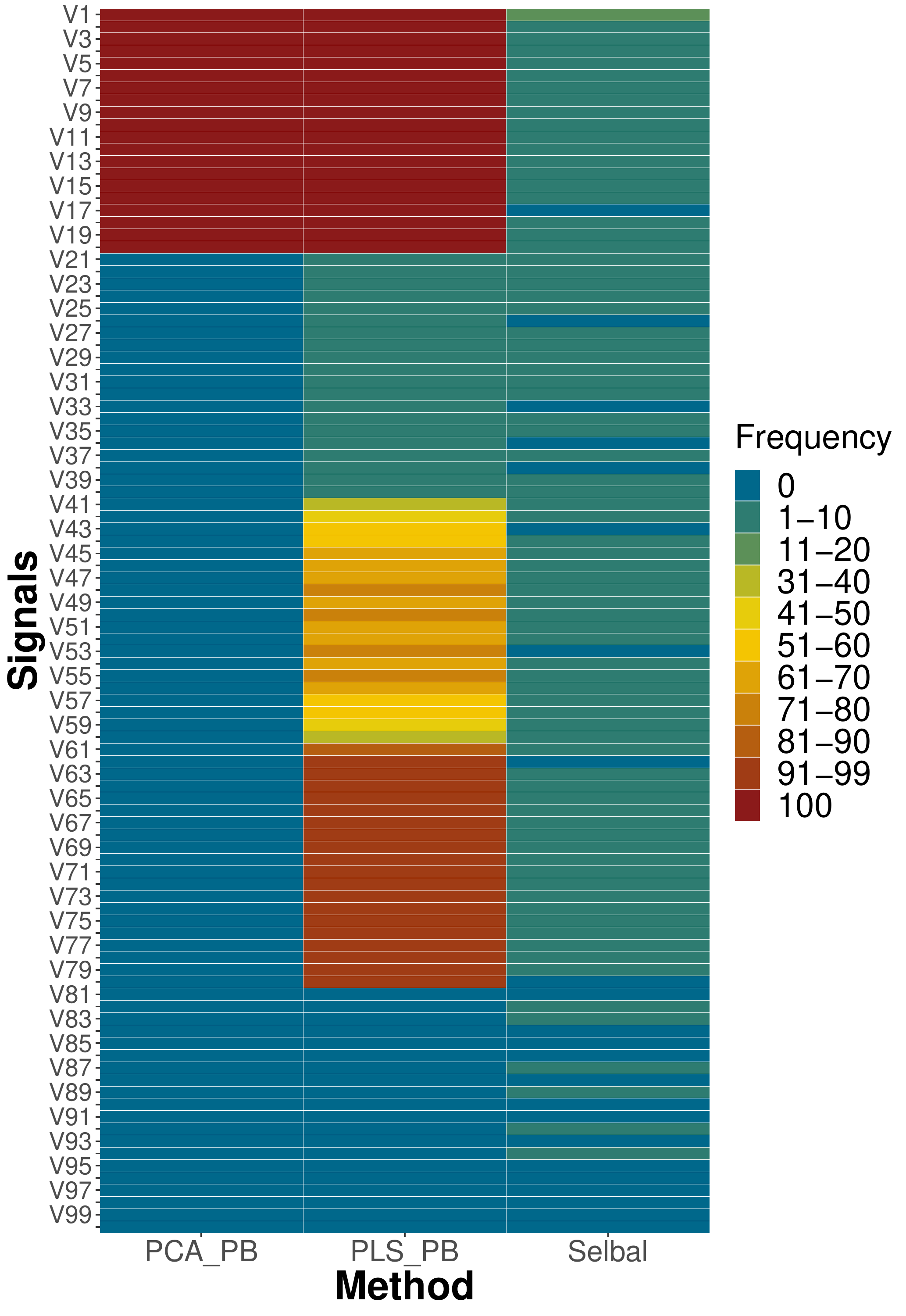}
	}
	\subfigure[Four different-sized blocks structure]{
		\label{subfig:lab3b}
		\includegraphics[width=0.3\textwidth]{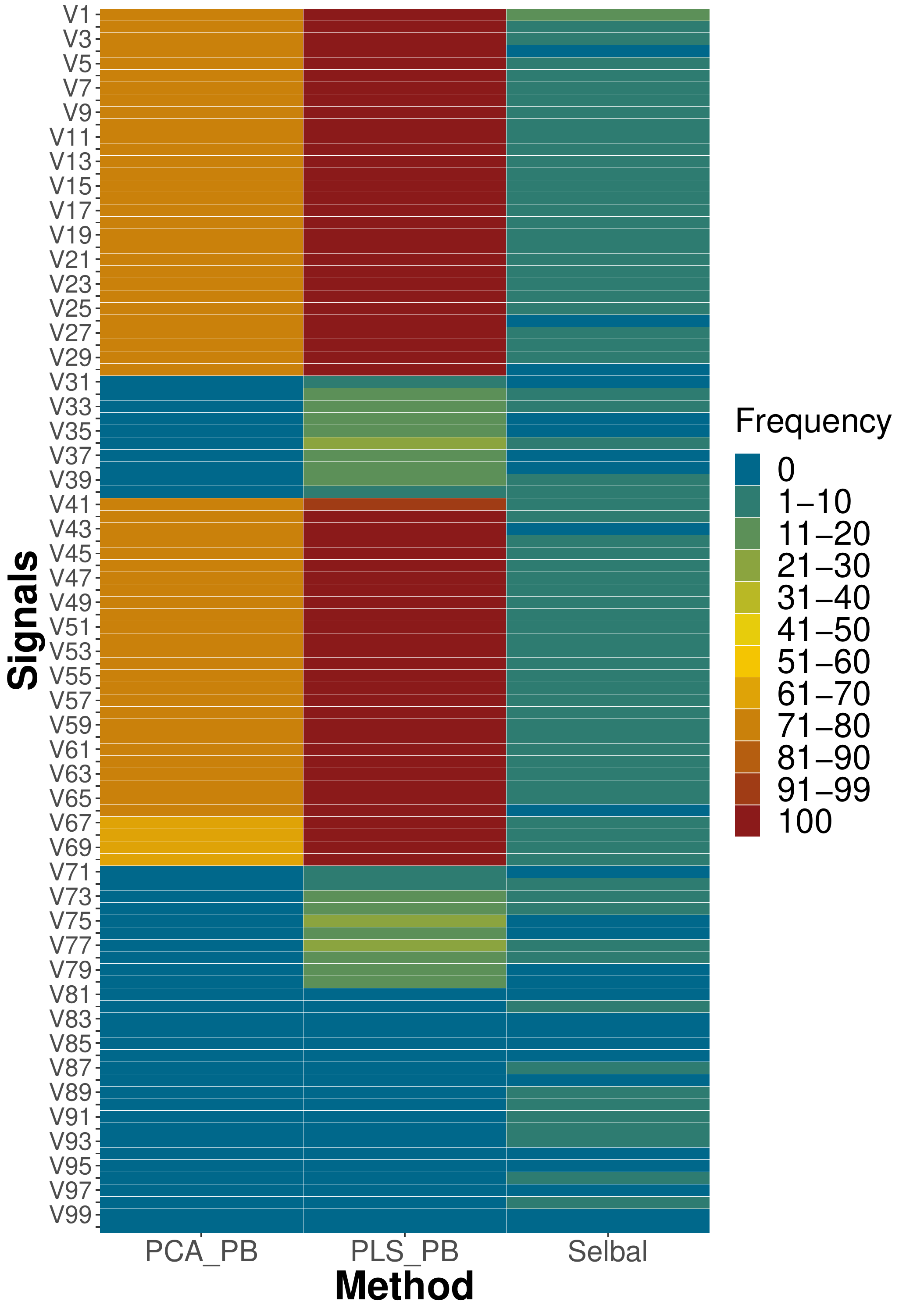}
	}
	
	\caption{Simulation: ability of the first balance to capture the data structure in the three simulation scenarios using the PCA-PB, PLS-PB and selbal algorithm.}
	\label{fig:SimRunsHeat}
	
\end{figure}

For each balance-based algorithm (PCA-PB, PLS-PB and selbal), we evaluate its ability to correctly identify the biomarkers. For each simulation run, the algorithms are applied to determine 
the first PB (PCA-PB and PLS-PB) and the optimal single balance (selbal). Then, for each signal, we record whether it is included in such balances or not. The heatmaps in Figure \ref{fig:SimRunsHeat} show the number of times a signal appears in the selected balance for each method, across the simulation runs. The selbal algorithm is unable to correctly distinguish marker from noise signals. In contrast, PLS-PB correctly identifies most of the markers. It even correctly detects signals in the milder marker blocks in most cases as observed in respectively Figure \ref{subfig:lab2b} and \ref{subfig:lab3b}. While PCA-PB performs similar to PLS-PB in the one-block scenario, it is only able to detect the most evident blocks of markers as observed in respectively Figure \ref{subfig:lab2b} and \ref{subfig:lab3b}.

%%%%%%%%%%%%%%%%%%%%%%%%%%%%%%%%%%%%%%%%%%%%%%%%%%%%%%%%%%%%%%%%%%%%%%%%
\section{Applications}    \label{RealDat}
%%%%%%%%%%%%%%%%%%%%%%%%%%%%%%%%%%%%%%%%%%%%%%%%%%%%%%%%%%%%%%%%%%%%%%%%

% R script: PBs_vzorovy_skript_Niky_data_CV.R (opt. number of bal)
% R script: PBs_sim_podle_NMR.R  (comparison of loadings and PBs)
% Heatmaps in scripts: PBs_sim_podle_NMR.R, PBs_sim_data_PLS_vs_PBs.R

We here demonstrate the application of the PLS-PB approach on two real-world data sets. Firstly, we consider a regression problem and then a classification task by simply accommodating the binary response into the PLS model.

\subsection{NMR Data Set}   \label{NMR}
We use the data set from \cite{stefelova2021weighted} consisting of high-throughput spectral profiles obtained by nuclear magnetic resonance (NMR). The data set involves a 127-part compositional predictor (metabolite signals; also called integrals) measured in $n = 211$ rumen fluid samples from cattle. This was collected along with individual measurements of animal methane yield ($\textrm{CH}_{4}$ in grams per kilogram of dry matter intake) which plays the role of a continuous response variable that we aim to model in terms of the metabolite composition. 

We apply the proposed PLS-PB method and analyze its ability to predict the response. Considering a varying number of PBs, we aim to detect an optimum that combines sensible prediction accuracy with preferably a small number of PB. Moreover, we investigate whether the PBs reflect the structure of PLS loadings while simplifying the interpretation.

\subsubsection{Optimal Number of Principal Balances}  \label{OptNbal}
Similar to the simulation study (Section \ref{Sim}), we compute PLS and PCA-PBs as well as standard PLS and compare prediction performance for models including several PBs (ranging from the first PB or PLS loading up to all $D-1$ = $126$ of them). Again, RMSEP is the measure used for comparison and its values are estimates based on $5$-fold CV and averaged over $100$ runs. 

For each number of latent components (either PB or ordinary loadings), the mean and standard deviation of the RMSEP values were computed across the 100 runs. Then, given the lowest mean RMSEP, the model using the fewest number of balances within one standard error from such a minimum is chosen (one standard error rule; see \citealp{friedman2001elements}). Accordingly, the most parsimonious model amongst those of best prediction performance is selected.

\begin{figure}[t]
	\centering
	\subfigure[Results for all latent components]{
		\label{subfig:a}
		\includegraphics[width=0.47\textwidth]{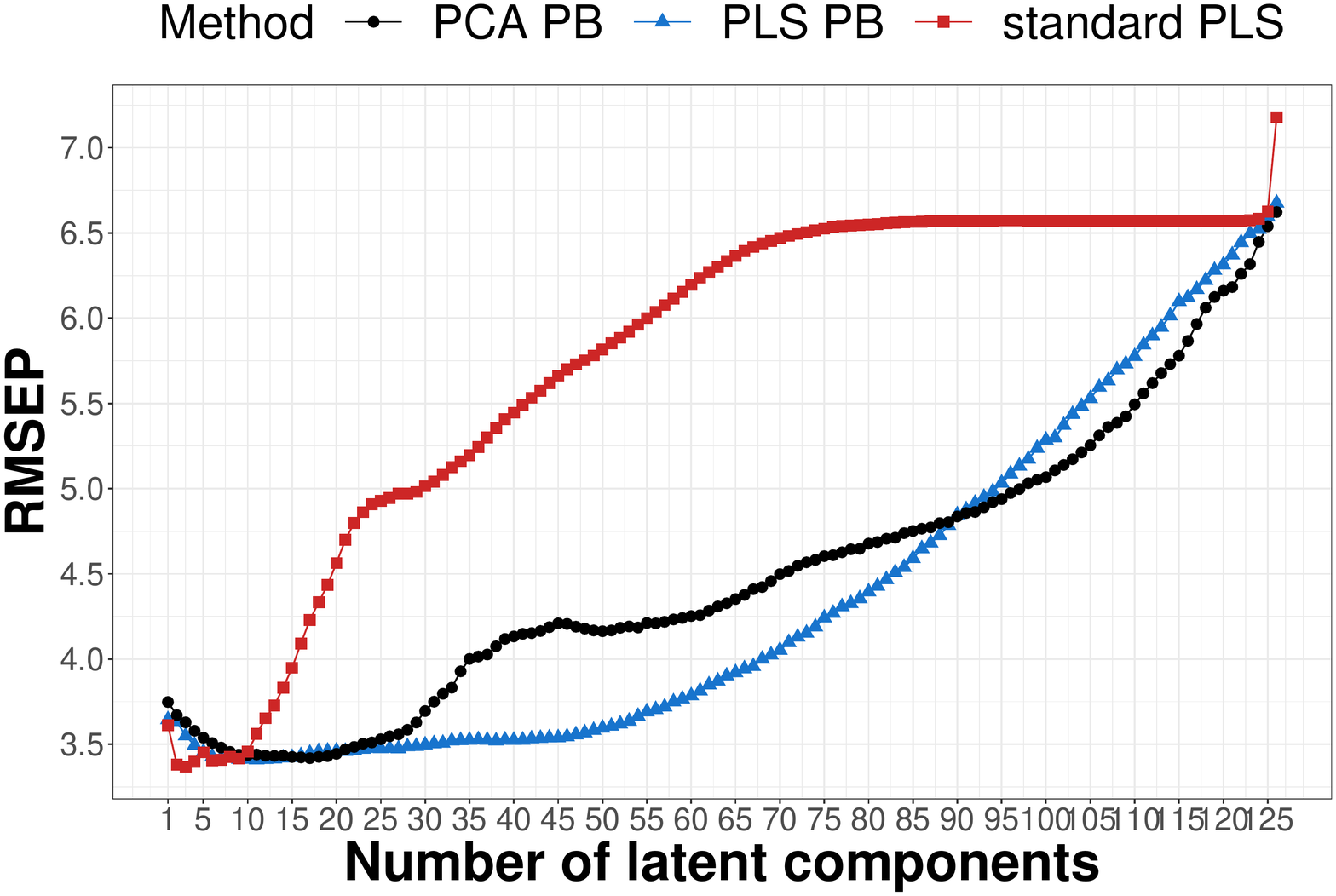}} 
	\subfigure[Results for up to $10$ latent components]{
		\label{subfig:b}
		\includegraphics[width=0.47\textwidth]{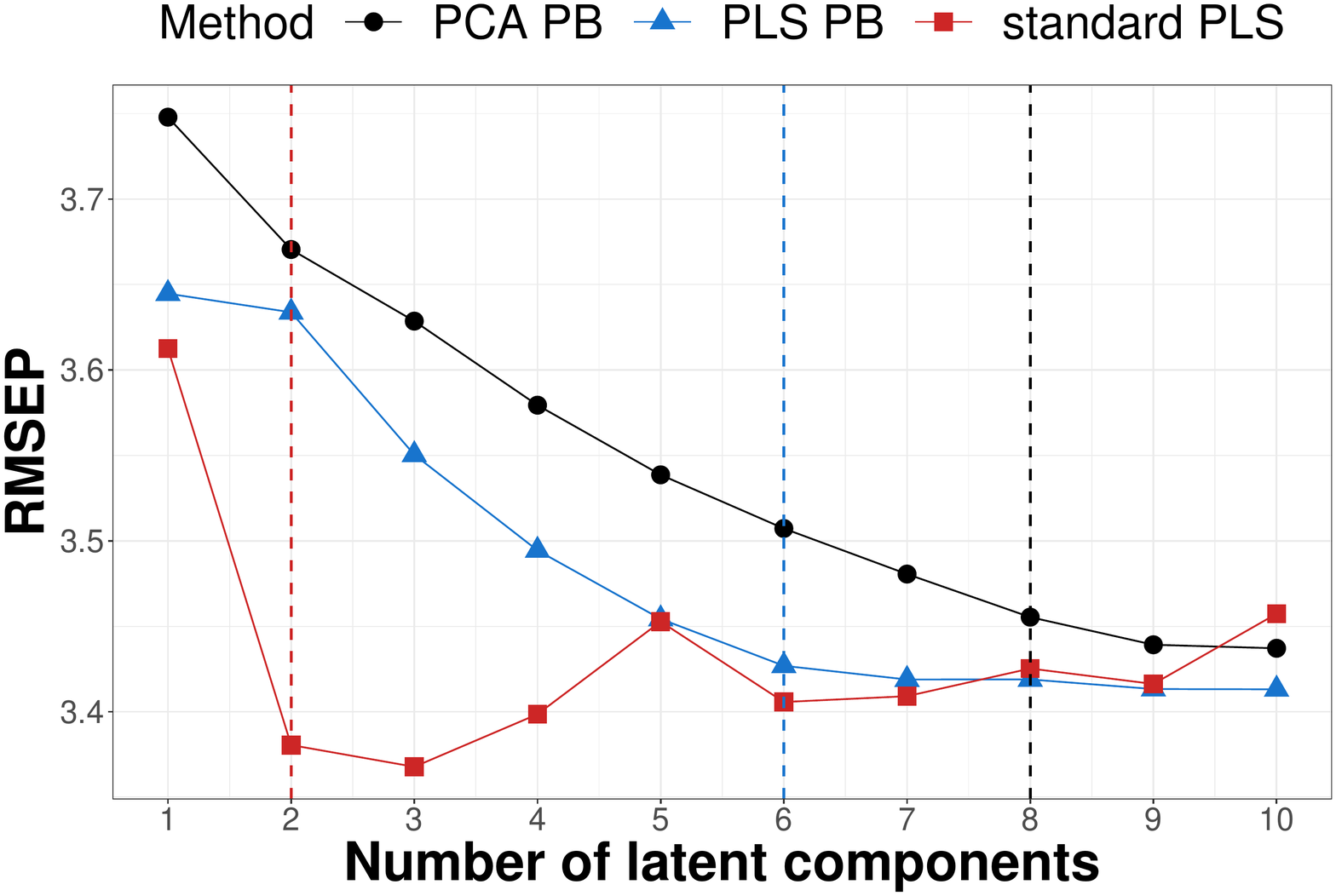}}
	\caption{Prediction performance of PLS-PB (blue), PCA-PB (black) and standard PLS (red) methods on NMR data set for different choices of latent components used (either PB or ordinary loadings). Optimal number determined according to one standard error rule from minimum CV RMSEP are indicated by vertical dashed lines for each method.}
	\label{fig:NMR_all}
\end{figure}

The results based on PLS-PB (blue) and PCA-PB (black), together with the results of standard PLS (red), are shown in Figure \ref{subfig:a}. We can observe that PLS-PB outperforms PCA-PB for most part of the range of PBs. Unlike in the previous simulation study, the effect of a possible overfitting issue can be observed  here as the RMSEP increases with the number of PBs. Also, as expected, the RMSEP values coincide (disregarding the minimal numerical difference) again for both approaches at the maximum number of PBs. Similarly to the simulation study in Section \ref{Sim}, standard PLS outperforms PLS-PB for the lowest numbers of latent components. However, weaker prediction performance of PLS-PB is compensated by the interpretability, which can be later seen in Figure \ref{fig:NMRbal}. Moreover, it can be clearly seen in Figure \ref{subfig:a} that standard PLS performs worse with increasing the number of loadings considered.

It can be observed in Figure \ref{subfig:a} that the RMSEP for %both PLS and PCA PBs 
all three methods decreases rapidly at the beginning of the range as PBs (or ordinary loadings) are aggregated. Figure \ref{subfig:b} zooms in on the results for the range of the first ten  latent components. While the smallest RMSEP for PLS-PB occurs for a model consisting of $11$ PBs,  a more parsimonious model using just $6$ PBs provides comparable performance according to the one standard error rule (marked by a blue vertical dashed line in Figure \ref{subfig:b}). For PCA-PB, the minimum occurs for a model containing the first $17$ PBs, while the model with $8$ PBs lies within a one standard error of this minimum (marked by a black vertical dashed line in Figure \ref{subfig:b}). For standard PLS, the minimum was reached for $3$ loadings and the optimal model determined by one standard error is the one with $2$ loadings (red vertical dashed line in Figure \ref{subfig:b}).
The prediction performance was also assessed using the selbal algorithm. In this case, the resulting RMSEP is $3.227$, much lower than for the other three methods. However, it was shown in Section \ref{Sim} that selbal has a rather poorer ability to capture the structure of the data compared to PLS-PB, which can be seen in Figure \ref{NMR:c}.

It is important to note that even if the PLS-PB approach does not outperform the PCA-PB approach, it is by construction expected to provide a more interpretable structure of PB, as these are tailored to maximize association with the response variable. The next section discusses the interpretative advantage of the PLS-PB approach.

\subsubsection{Comparison of PLS-PB to PLS Loadings} \label{SimRes}
We compare the PLS-PBs to PLS loadings to examine whether the former suitably reflect the latter (in terms of signs of coefficients) and, at the same time, facilitate interpretation.

\begin{figure}
	\centering
	\subfigure[PLS loadings]{
		\label{NMR:a}
		\includegraphics[width=0.31\textwidth]{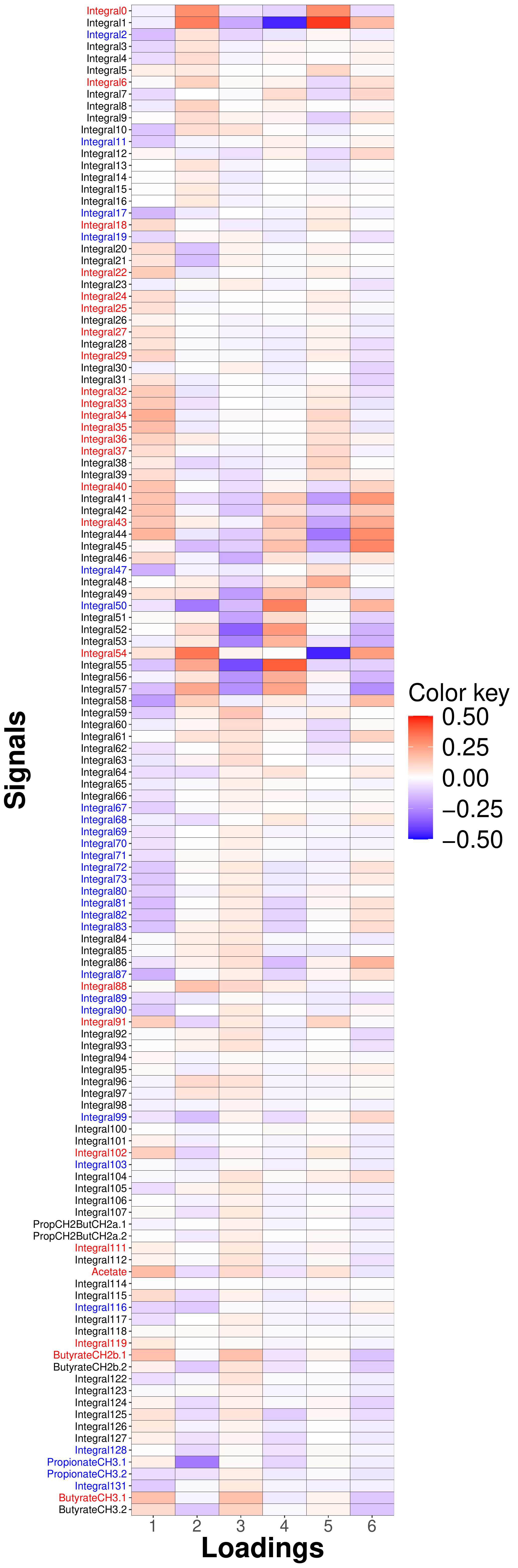}} 
	\subfigure[PLS-PB]{
		\label{NMR:b}
		\includegraphics[width=0.31\textwidth]{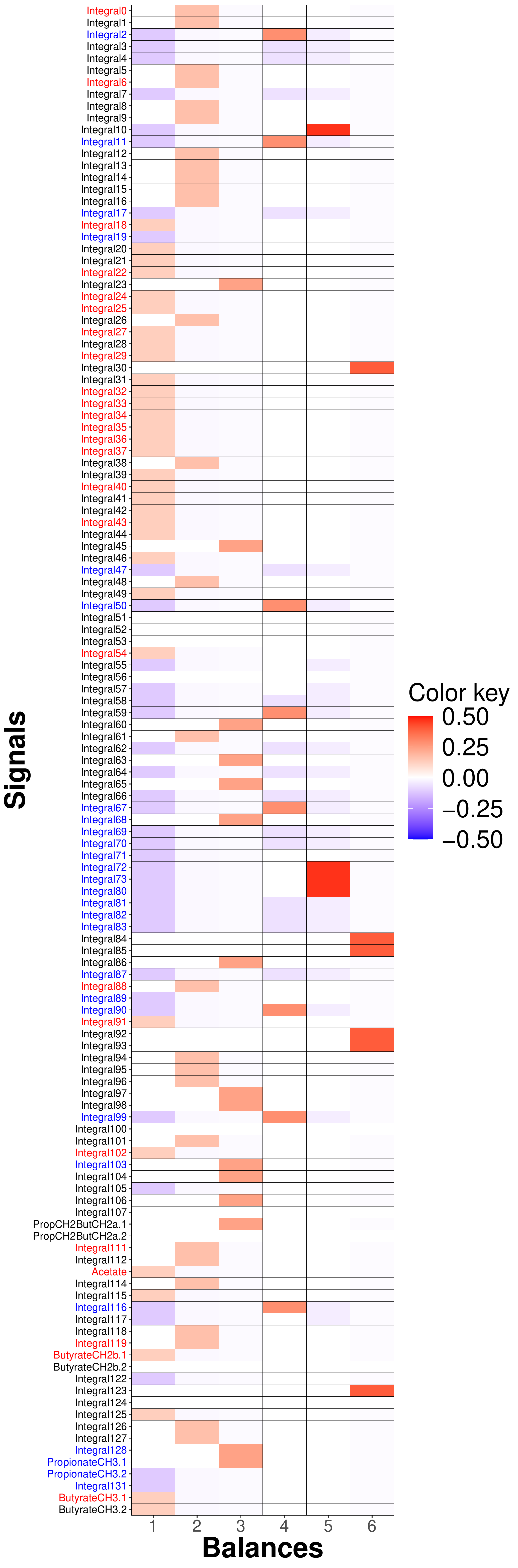}} 
	\subfigure[selbal balance]{
		\label{NMR:c}
		\includegraphics[width=0.31\textwidth]{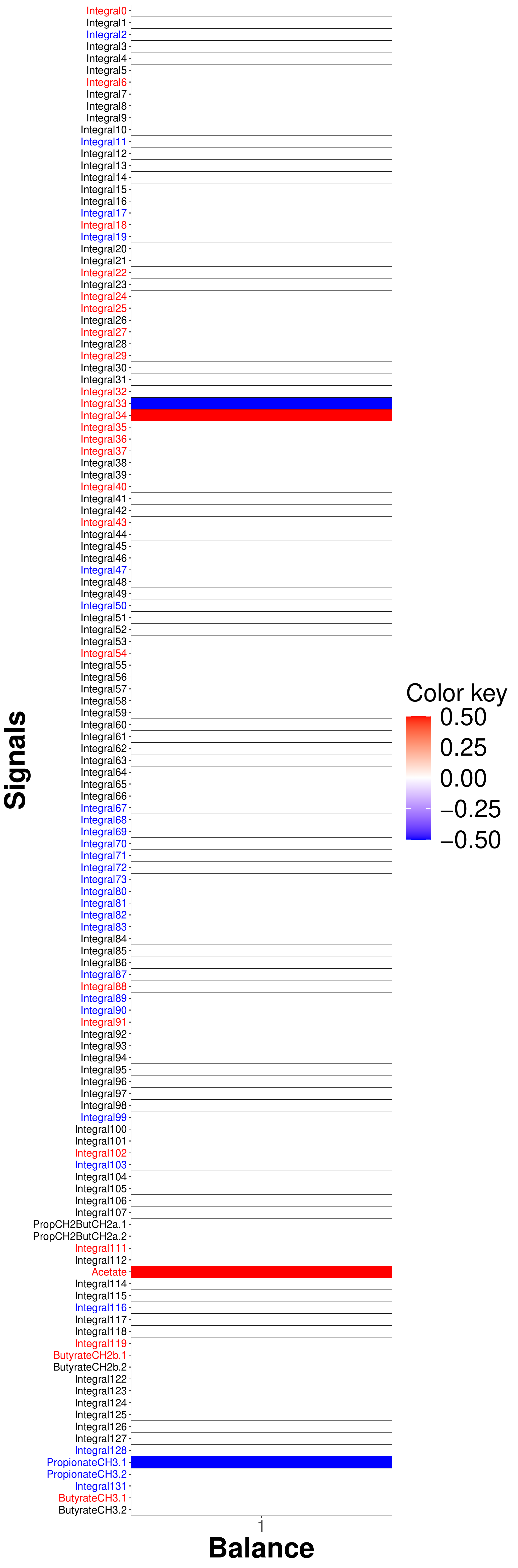}}
	\caption{Comparison of the first $6$ PLS loadings, $6$ PLS-PB and a selbal balance from the NMR data set. Red and blue labels are used to highlight markers identified in previous studies. Red (blue) colored text is used for markers having a positive (negative) relationship with the response variable.}
	\label{fig:NMRbal}
\end{figure}

In Figure \ref{fig:NMRbal} we display their values for the first $6$ PBs, the optimal number determined in the previous section. The first PLS-PB reproduces the structure of signs of the coefficient values observed in the first PLS loading vector well, highlighting biologically meaningful markers identified in~\cite{stefelova2021weighted}, related to methane yield. 
For example, very distinct groups of identified markers are the group from Integral$32$ to Integral$37$ and a group from Integral$67$ to Integral$83$ (with Integral$68$ being picked in the third balance). These markers are colored red and blue in Figure \ref{fig:NMRbal}. Markers colored in red are those having a positive relationship with the response variable, whereas blue-colored markers are those that have a negative relationship with the response variable. The other PLS-PB, whose structure does not necessarily coincide with the structure of the respective PLS loadings due to the orthogonality constraint, capture some other patterns, related to both marker and non-marker variables. Moreover, PLS loadings also provide misleading information, as in the third loading there is a group of signals (Integral$49$ to Integral$57$) which is highlighted, but it is not biologically meaningful~\cite{stefelova2021weighted}. On the contrary, PLS-PB provides a neater and more parsimonious view.

\subsection{Metabolomic Data Set}
We now consider a metabolomic data set consisting of $n = 46$ observations and $D = 209$ metabolites, thus representing the common high-dimensional setting with $n < D$. The response variable is in this case dichotomous and states cancerous ($y_{i} = 1$) or healthy ($y_{i} = 0$) tissues, having $23$ patients suffering from lung cancer and other $23$ being healthy \citep{cifkova2022lipidomic}. The aim is to enable classification of tissues, and in particular, to reveal pathobiochemical changes of the disease. The samples were analysed by a targeted metabolomic method based on HILIC liquid chromatography coupled with triple quadrupole mass spectrometry. This method allows to detect altogether 350 metabolites in different biofluids, tissues and cells and covers main metabolic pathways.

We applied the PLS-PB and PLS-PCA methods and assessed their relative performance by 5-fold CV over the range of possible numbers of PB as detailed previously. The RMSEP was replaced by the misclassification error, defined as
\begin{equation}
	\label{miscl}
	\mathrm{ME}=\frac{1}{n}\sum_{i=1}^n I(\hat{y}_i\neq y_i),
\end{equation}
where $y_i$ denotes the group number of the $i$th object, $\hat{y}_i$ is the estimated group number, and the index function $I$ gives 1 if the group numbers are not the same and 0 otherwise.

\begin{figure}[t]
	\centering
	\includegraphics[width=0.6\textwidth]{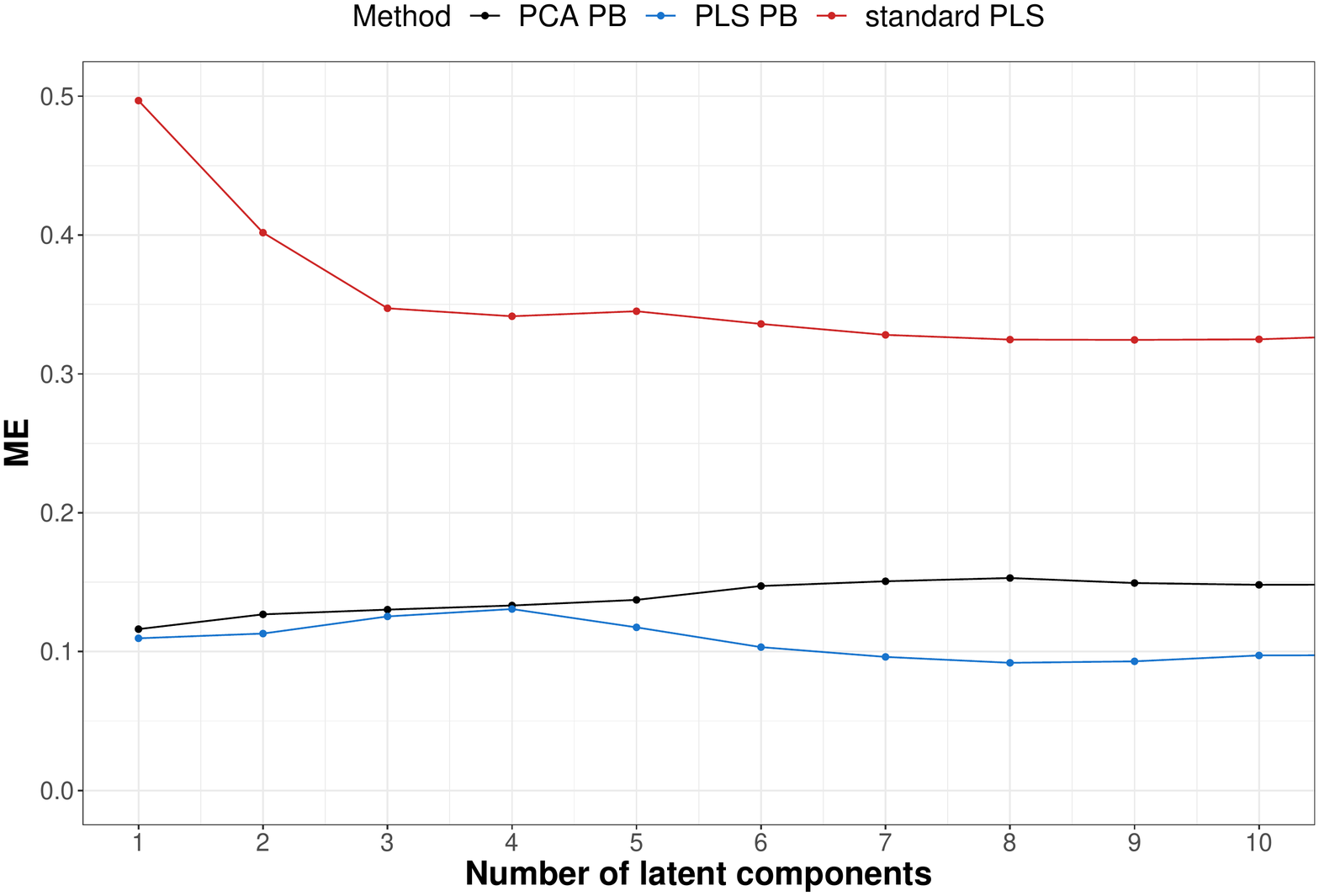}.
	\caption{Prediction performance of PLS-PB (blue), PCA-PB (black) and standard PLS (red) methods on metabolomic data set across the first 10 latent components.}
	\label{fig:MetabCV}
\end{figure}

Figure \ref{fig:MetabCV} displays the results for the first $10$ latent components. Similar to the NMR data study, PLS-PB generally outperforms PCA-PB. Even though the numerical difference in ME is not dramatic, it can be seen that the ME of PLS-PB slightly drops, whereas the ME of PCA-PB rather levels off. On the other hand, the performance of standard PLS is considerably worse than the other methods. Their performance was again compared to the result of selbal, for which the ME was $0.343$. Selbal thus shows worse performance than PLS-PB and PCA-PB. Moreover, this latter does not recover the structure of markers very well.

\begin{figure}
	\centering
	\subfigure[PLS loadings]{\includegraphics[width=0.4\textwidth]{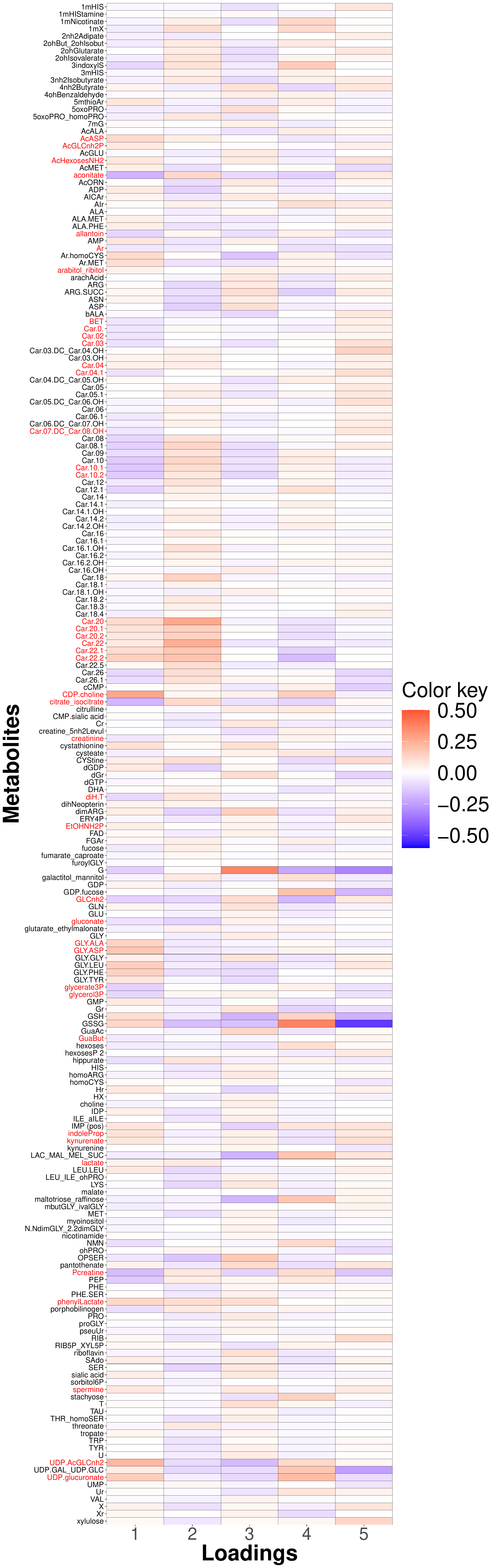}} 
	\subfigure[PLS-PB]{\includegraphics[width=0.4\textwidth]{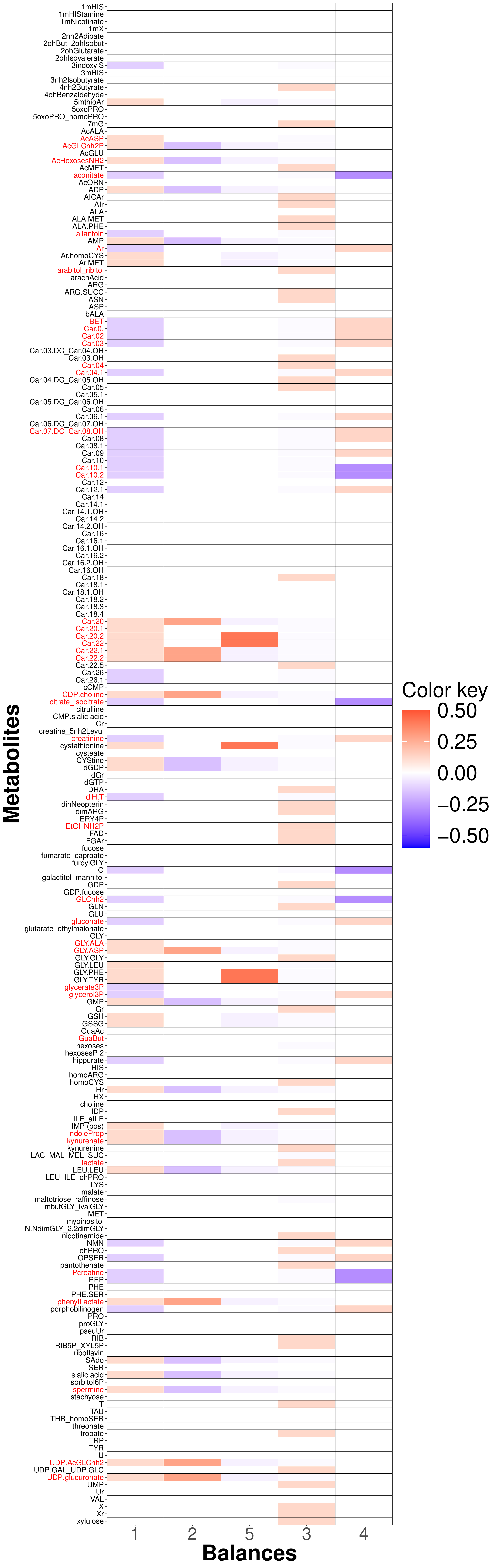}} 
	\caption{Comparison of the first five PLS loadings and PLS-PB. Red colored text is used to highlight metabolites which were marked as significant using $p$-values after Bonferroni correction.}
	\label{fig:DataDavid}
\end{figure}

Figure \ref{fig:DataDavid} displays PLS loadings and PLS-PB for the first five balances. Again, PLS-PBs appear to be less noisy than the PLS loadings counterparts. The latter puts an unnecessarily large emphasis on absolute differences. PLS-PBs, in contrast, are easier to navigate in the outcome matrix and show more agreement with univariate statistical analysis \citep{cifkova2022lipidomic}. 
We can see general trends in decreasing short and medium chain (Car.0 - Car.12) compared to increased very long chain (Car.20 - Car.22) acyl carnitines, which are closely metabolically connected. Furthermore, selected groups of metabolites such as glycine dipeptides (GLY.ALA - GLY.TYR) and pyrimidine nucleotides (UDP.glucuronate, UDP.AcGlcNH2 and CDP.choline) show systematic trends.
The PLS-PB approach in fact splits acylcarnitines into two separate groups, which could be then subject of future research.

%%%%%%%%%%%%%%%%%%%%%%%%%%%%%%%%%%%%%%%%%%%%%%%%%%%%%%%%%%%%%%%%%%%%%%%%
\section{Conclusions}    \label{end}
%%%%%%%%%%%%%%%%%%%%%%%%%%%%%%%%%%%%%%%%%%%%%%%%%%%%%%%%%%%%%%%%%%%%%%%%
This manuscript introduces a new procedure to construct PBs within a log-ratio analysis framework for high-dimensional CoDa. We extend previous work in PBs by exploiting PLS as a dimension reduction tool that accounts for the relationship between a response variable of interest and a high-dimensional composition playing the role of predictor. The algorithm determines $D-1$ data-driven PLS-PBs that maximize their covariance with the response variable.

The proposal is applicable to both regression and classification problems and our numerical experiments firstly demonstrate that the resulting PLS-PBs provide a simplified structure of PLS loadings and outperform the original PCA-PB in terms of prediction performance. Secondly, when compared with the recently proposed selbal algorithm, which targets the same goal as PLS-PB, it is shown that although the selbal method may perform better in terms of prediction, it shows poorer ability to capture the data structure. Finally, PLS-PBs simplify the structure and enhance the interpretation of the results when compared with standard PLS.  The method is further demonstrated on two real data sets regarding regression analysis with NMR spectral data and a classification task with metabolomic data. In both cases, the usefulness of the PLS-PB approach for variable selection and biomarker discovery is illustrated.

Building on the PLS-PB framework presented here, possibilities for further developments include its robustification to manage the potential influence of outlying samples in the results or the ability to deal with sparse data.

%%%%%%%%%%%%%%%%%%%%%%%%%%%%%%%%%%%%%%%%%%%%%%%%%%%%%%%%%%%%%%%%%%%%%%%%
\section*{Acknowledgments}

JPA and KH gratefully acknowledge the support by the project PID2021-123833OB-I00 supported by the Spanish Ministry of Science and Innovation (MCIN/AEI/10.13039/501100011033) and ERDF A way of making Europe; KH and PF were supported by the Czech Science Foundation, Project 22-15684L, and by the Austrian Science Foundation, Project I 5799-N, respectively; VN and KH were supported by IGA\_PrF\_2022\_008 Mathematical models.

%%%%%%%%%%%%%%%%%%%%%%%%%%%%%%%%%%%%%%%%%%%%%%%%%%%%%%%%%%%%%%%%%%%%%%%%
\bibliographystyle{apalike}
\bibliography{Main}
\nocite{*}

%%%%%%%%%%%%%%%%%%%%%%%%%%%%%%%%%%%%%%%%%%%%%%%%%%%%%%%%%%%%%%%%%%%%%%%%
%%% APPENDIX
\newpage
\appendix
\section{Artificial Settings for Comparison: Simulation Design} \label{appA}

Compositions were simulated using so-called pivot coordinates, an instance of olr coordinates \citep{FiHr2011}, and assuming multivariate normality. In general, pivot coordinates are defined as 
\begin{eqnarray*}
	z_{j}^{(l)} &=& \sqrt{\frac{D-j}{D-j+1}} \textrm{ln} \frac{x_{j}^{(l)}}{\sqrt{\prod_{k = j+1}^{D}x_{k}^{(l)}}} \nonumber \\ &=& \frac{1}{\sqrt{(D-j+1)(D-j)}} \Big[ \textrm{ln} \Big( \frac{x_{j}^{(l)}}{x_{j+1}^{(l)}} + \dots +  \frac{x_{j}^{(l)}}{x_{D}^{(l)}}\Big) \Big] , l = 1,\dots,D, \: j = 1,\dots,D-1,
\end{eqnarray*}
where $\mathbf{x}^{(l)} = (x_{1}^{(l)}, \dots, x_{D}^{(l)})^\top$ is a rearranged composition $\mathbf{x}$ having the $l$-th part on the first position. It follows that via pivot coordinates, the relative information about the $l$-th part is captured by the first coordinate, which is advantageously used here for setting up our example. That is, it holds $z_{1}^{(l)}=\sqrt{\frac{D}{D-1}}\;\textrm{clr}_{l}(\mathbf{x})$ for $l = 1,\dots,D$.

The steps to generate data (proposed in \cite{stefelova2021weighted}) can be summarized as follows:  

\begin{itemize}
	\item Firstly, generate pivot coordinates $\mathbf{z}_{i} = (z_{i,1},\dots , \allowbreak z_{i,D-1})^\top$ from a multivariate normal distribution  $N_{D-1}(\mathbf{0},\mathbf{\Sigma})$, $i = 1,\dots,n$.
	The elements (with $i,j = 1, \dots, D-1$) in the covariance matrix $\mathbf{\Sigma}$ are  equal to 
	\begin{equation}
		\label{simcov}
		\sigma_{i,j} = 
		\begin{cases} 
			2 & \text{if } i = j \leq 2r \\
			1 & \text{if } i = j > 2r \\
			0.5 \times(-1)^{i+j} & \text{if } i \neq j, i,j \leq 2r \\
			0 & \text{otherwise,}
		\end{cases}
	\end{equation}
	where, in our setting, the first $2r$ compositional parts thus act as markers in the block of the first example, as set by the structure of the covariance matrix. To be more specific, first $r$ odd principal balances are modeled to have positive covariance between each pair. Similarly, first $r$ even principal balances are then modeled to have negative covariance between each pair.
	
	\item Secondly, to obtain matrix $\mathbf{X}$, pivot coordinates need to be back-transformed:
	$$
	\mathbf{x}_{i} = \textrm{ilr}^{-1}(\mathbf{z}_{i}) = (x_{i,1},\dots,x_{i,D})^\top, i = 1,\dots,n.
	$$
	
	\item Finally, the response variable then results from $$ y_{i} = \beta_{1}z_{i,1} - \beta_{2}z_{i,2} +  \dots + \beta_{2r-1}z_{i,2r-1} - \beta_{2r}z_{i,2r} + \varepsilon_{i},$$
	where $\varepsilon_{i} \sim N(0,1)$, $i = 1,\dots,n$, and $\beta_{j} \sim U(0.1,1)$, $j = 1,\dots,2r$.
	The first $2r$ pivot coordinates are used as predictor variables. For the first $r$ odd pivot coordinates, we take positive regression coefficients; for the first $r$ even pivot coordinates we take negative ones. 
\end{itemize}

\end{document}